\documentclass[%
 reprint,
 amsmath,amssymb,
 aps,
]{revtex4-2}

\usepackage{graphicx}
\usepackage{dcolumn}
\usepackage{bm}
\usepackage{hyperref}
\usepackage{xcolor}

\begin{document}

\preprint{APS/123-QED}

\title{Discrete-phase-randomized mode-pairing quantum key distribution}

\author{Yuewei Xu$^{1}$}
\author{Zeyang Lu$^{1}$}
\author{Chan Li$^{1}$}
\author{Jian Long$^{1}$}
\email{longjian@ecust.edu.cn}
\author{Zhu Cao$^{2,3}$}%
\email{caozhu@tongji.edu.cn}
\affiliation{$^{1}$Key Laboratory of Smart Manufacturing in Energy Chemical Process,
Ministry of Education, East China University of Science and Technology,
Shanghai, 200237, China \\
$^{2}$College of Electronics and Information Engineering,
Tongji University, Shanghai 201804, China \\
$^{3}$Shanghai Research Institute for Intelligent Autonomous Systems,
Tongji University, Shanghai 201210, China}


\begin{abstract}
Mode-pairing quantum key distribution (MP-QKD) protocol achieves performance beyond the repeaterless rate-transmittance bound and exhibits excellent practicality by avoiding the requirement for difficult global phase locking. However, the source side of MP-QKD still relies on the assumption of continuous phase randomization, an experimentally infeasible requirement in practice. Therefore, the practical security of the protocol cannot be fully guaranteed. In this work, we propose a discrete-phase-randomized mode-pairing quantum key distribution (DPR-MP-QKD) protocol and analyze the basis-dependence of the source side. Then, we introduce a concrete discrete version of the decoy state method that ensures the security of the DPR-MP-QKD protocol. Finally, simulation results indicate that as the number of discrete phases increases, the key rate performance of DPR-MP-QKD progressively approaches that of the continuous case, with convergence achieved at approximately 14 discrete phases. Moreover, our approach drastically lowers the demand for randomness. While conventional continuous phase randomization demands an unlimited supply of random bits, we show that merely a few bits (e.g., 4) are adequate.
\end{abstract}


\maketitle


\section{\label{sec:level1}Introduction}

Quantum key distribution (QKD) provides a foundational method for achieving information-theoretic secure communication between two distant users \cite{bennett2014quantum}, positioning it as a cornerstone for future quantum communication networks \cite{chen2021integrated}. Its prominent advantage lies in advanced technological maturity, having evolved from theoretical proof to experimental demonstration and early commercial deployment \cite{pirandola2020advances}. However, practical deployment faces two major challenges: practical security vulnerabilities \cite{lydersen2010hacking, xu2020secure} and key rate performance limitations \cite{takeoka2014fundamental, pirandola2017fundamental}.

In terms of security, discrepancies between realistic devices and their idealized models introduce potential loopholes. The security of QKD systems can be compromised by various quantum attacks that exploit practical imperfections in their physical components \cite{xu2020secure}. A typical QKD setup consists of three main elements: source, quantum channel, and measurement apparatus. Channel security has been rigorously established within existing QKD security proofs \cite{lo1999unconditional, shor2000simple, koashi2009simple}, while the source, being relatively simple \cite{lo2007security}, can be well modeled. Nevertheless, the measurement devices, being complex and susceptible to external manipulation, constitute a major security weakness. To address all detection side loopholes, measurement-device-independent QKD (MDI-QKD) was developed \cite{lo2012measurement}. Subsequently, phase-encoded variants of the MDI scheme also emerged, where the key information is encoded using two optical time bins \cite{tamaki2012phase, ma2012alternative}. Compared to conventional QKD, the key advantage of MDI-QKD lies in its independence from the physical implementation of the detection side. Since neither of the communicating parties acts as a receiver of quantum signals during key distribution, an attacker is effectively prevented from compromising the system by targeting the measurement apparatus. This approach has been validated through various experimental demonstrations and has been extended to network environments \cite{liu2013experimental, ferreira2013proof, tang2014experimental, woodward2021gigahertz, tang2016measurement}.

Regarding performance, optical channel loss severely limits the key generation rate. At present, the performance of point-to-point QKD is limited by the repeaterless rate-transmittance bound (or PLOB bound) \cite{takeoka2014fundamental, pirandola2017fundamental}, and the key rate can only increase linearly with channel transmittance. Although quantum repeaters offer a theoretical solution \cite{zukowski1993event, briegel1998quantum, azuma2015all}, their near-term realization still faces significant challenges. Remarkably, twin-field QKD (TF-QKD) \cite{lucamarini2018overcoming}, based on single-photon interference, successfully overcomes this PLOB bound, achieving a key rate that scales with the square root of transmittance. Later, phase-encoded variant (Phase-Matching QKD) and intensity-encoded variant (Sending-or-Not-Sending TF-QKD) of the TF-QKD have been developed \cite{lucamarini2018overcoming, ma2018phase, lin2018simple, wang2018twin}. Unfortunately, due to the lack of a global phase reference in such TF-type protocols, the implementation of global phase locking is required, which poses significant challenges for practical realization.

A recently proposed MDI-type protocol, mode-pairing quantum key distribution (MP-QKD), demonstrates unique flexibility and enhanced key rate performance \cite{zeng2022mode}. Specifically, MP-QKD encodes key information in the relative phase of paired pulses, enabling it to achieve the same encoding efficiency as TF-type protocols without global phase locking while maintaining stable interference under mild local phase fluctuations. The performance of this protocol has been verified experimentally \cite{zhu2023experimental}. However, its source side security requires further refinement. In particular, idealized assumptions such as continuous phase randomization are difficult to satisfy perfectly in experiments, potentially introducing security threats \cite{sun2012partially, tang2013source}.

In this work, inspired by previous discrete-phase-randomization approaches developed for BB84 and MDI-QKD to address source side phase randomization loopholes \cite{cao2015discrete, cao2020discrete}, we propose a discrete-phase-randomized mode-pairing quantum key distribution (DPR-MP-QKD) protocol. Then we derive the basis-dependence formula for the source and prove that only the pseudo single-photon states within it can contribute positively to the key rate. Furthermore, we present a concrete discrete-phase-randomized decoy state method that ensures the security of the DPR-MP-QKD protocol. Finally, the simulation results indicate that as the number of discrete phases increases, the key rate performance of DPR-MP-QKD progressively approaches that of the continuous case, with convergence achieved at approximately 14 discrete phases. Compared to conventional continuous phase randomization which consumes infinite random bits, our scheme significantly reduces this requirement to as few as 4 bits. Moreover, even with a fixed number of discrete phases, the key rate can still surpass the PLOB bound as the pairing interval increases.

The structure of the remainder of this paper is organized as follows. In Sec.~\ref{sec2}, we introduce discrete-phase-randomized coherent-state sources for both single-mode and dual-mode encoding scenarios. We then analyze the security components of DPR-MP-QKD through basis dependence, derive its key rate expression, and finally present the DPR-MP-QKD protocol. In Sec.~\ref{sec3}, we analyze the security of the DPR-MP-QKD by using the discrete-phase-randomized decoy state method. Numerical simulation results for DPR-MP-QKD are provided in Sec.~\ref{sec4}. We conclude the paper and discuss future research directions in Sec.~\ref{sec5}.







\section{\label{sec2}Discrete-phase-randomized mode-pairing quantum key distribution}
In this section, we first elucidate the differences between discrete-phase-randomized light sources and the continuous case. Based on the number of optical modes encoded by random phases, we will analyze discrete-phase-randomized coherent state sources by categorizing them into single-optical-mode and dual-optical-mode scenarios. Subsequently, we will analyze the security components based on the basis-dependence. Finally, we will propose a discrete-phase-randomized mode-pairing quantum key distribution protocol.

In the case of continuous phase randomization, a coherent state $\left | \alpha \right \rangle$ is modulated by a random phase $\theta \in [0, 2\pi)$, resulting in the state $\left | \alpha e^{i\theta} \right \rangle$. Since a coherent state is a quantum superposition of Fock states, it possesses non-zero probability amplitudes not only for the single-photon state, but also for the vacuum and multi-photon states. When the phase of a coherent state is taken to be fully randomized, we can use the photon-number channel model to describe this source flaw \cite{lo2005decoy},
\begin{equation}
\frac{1}{2\pi}\int_{0}^{2\pi} \left | \alpha e^{i\theta} \right \rangle 
\left \langle \alpha e^{i\theta} \right | d\theta = \sum_{k=0}^{\infty } 
e^{-\left | \alpha \right |^2 } \frac{\left | \alpha \right |^{2k}}{k!} 
\left | k  \right \rangle \left \langle k \right |.
\label{cpr}
\end{equation}
Essentially, Eq. (\ref{cpr}) indicates that perfect phase randomization of a coherent state renders it equivalent to a statistical mixture of Fock states obeying a Poisson distribution with mean $|\alpha|^2$. In other words, continuous phase randomization leads to complete decoherence among the Fock states.

Under the condition of discrete phase randomization for coherent state light sources, the phase values are no longer infinite. For simplicity, the range $\left[0, 2\pi\right)$ can be uniformly discretized into $D$ phase values, with each phase selected randomly with equal probability $1/D$. The set of discrete phases is defined as
\begin{equation}
    \left \{ \phi_{n} = \frac{2\pi}{D}n \mid n\in 0, 1, 2, ..., D-1 \right \},
    \label{dp}
\end{equation}
where $D$ is an even integer that represents the number of discrete phases. In this scenario, we cannot decompose the light source into independent Fock states for subsequent security analysis. Later, we will examine separately the cases where random phases are encoded in a single optical mode and in dual orthogonal optical modes.

\subsection{\label{sec:som}{Single-optical mode case}}
Consider a coherent state $\left | \sqrt{\mu} e^{i \phi _{n}} \right \rangle$ on system $A$ with $ \phi_{n}$ coming from Eq. (\ref{dp}). By introducing a virtual ancilla qudit system $\widetilde{A}$ with dimension $d=D$ to record the discrete phase information, the joint state can be expressed as \cite{cao2015discrete}
\begin{equation}
    \left | \psi_{D}  \right \rangle _{\widetilde{A} A} = \frac{1}{\sqrt{D}} \sum_{n=0}^{D-1} 
    \left | n \right \rangle _{\widetilde{A} } 
    \left | \sqrt{\mu} e^{i \phi _{n}} \right \rangle _{A}.
\label{som}
\end{equation}
To obtain the photon number information about the emitted state by measuring the ancilla qudit system $\widetilde{A}$, we introduce a complementary orthogonal basis $\left \{ \left | k \right \rangle _{\widetilde{A}} \right \} _{k=0}^{D-1}$ in system $\widetilde{A} $ via the Fourier transform,
\begin{equation}
\begin{aligned}
    \left | k \right \rangle _{\widetilde{A}} & = \frac{1}{\sqrt{D}} \sum_{n=0}^{D-1} e^{i \frac{2 \pi}{D} nk} \left | n \right \rangle _{\widetilde{A}},\\
    \left | n \right \rangle _{\widetilde{A}} & = \frac{1}{\sqrt{D}} \sum_{k=0}^{D-1} e^{-i \frac{2 \pi}{D} nk} \left | k \right \rangle _{\widetilde{A}}.
\end{aligned}
\label{cb}
\end{equation}
As a result, we can reformulate the state in Eq.(2) as
\begin{equation}
\begin{aligned}
    \left | \psi_{D}  \right \rangle _{\widetilde{A}A} & = \frac{1}{\sqrt{D}} \sum_{n=0}^{D-1} 
    \left | n \right \rangle _{\widetilde{A}} 
    \left | \sqrt{\mu} e^{i \phi _{n}} \right \rangle _{A} \\
    & = \frac{1}{D} \sum_{n=0}^{D-1} \sum_{k=0}^{D-1} e^{-i \frac{2 \pi}{D} nk} 
    \left | k \right \rangle _{\widetilde{A}}
    \left | \sqrt{\mu} e^{i \phi _{n}} \right \rangle _{A} \\
    & = \sum_{k=0}^{D-1} \sqrt[]{P _{k}^{\mu}} 
    \left | k \right \rangle _{\widetilde{A}}
    \left | \lambda _{k}^{\mu} \right \rangle _{A},
\end{aligned}
\end{equation}
where the normalized state on the system $A$ is given by
\begin{equation}
\begin{aligned}
    \left | \lambda_{k}^{\mu} \right \rangle_{A} & = \frac{1}{D \sqrt[]{P _{k}^{\mu}}} \sum_{n=0}^{D-1} 
    e^{-i \frac{2 \pi}{D} nk} \left | \sqrt{\mu} e^{i \phi _{n}} \right \rangle_{A} \\
    & = \frac{1}{\sqrt[]{P _{k}^{\mu}}} e^{-\frac{\mu}{2}} \sum_{m=0}^{\infty}
    \frac{\sqrt[]{\mu}^{\left ( mD+k \right ) }}{\sqrt[]{\left ( mD+k \right ) ! } } 
    \left | mD+k  \right \rangle_{A},
\label{pks}
\end{aligned}
\end{equation}
and the corresponding probability of obtaining $\left | \lambda_{k}^{\mu} \right \rangle$ is
\begin{equation}
    P _{k}^{\mu} = e^{-\mu} \sum_{m=0}^{\infty}
    \frac{\mu ^{\left ( mD+k \right ) }}{\left ( mD+k \right ) ! }.
\label{dpr-pro}
\end{equation}

As can be seen from the above equation, the state $\left | \lambda_{k}^{\mu} \right \rangle$ is a superposition of Fock states with $k$ mod $D$. Meanwhile, as $D$ increases, state $\left | \lambda_{k}^{\mu} \right \rangle$ asymptotically approaches a Fock state. In the limit $D \to \infty$, it converges exactly to the Fock state $\left| k \right\rangle$, and the corresponding probability distribution becomes a Poisson distribution, $e^{-\mu}\frac{\mu ^{k}}{k!}$. Therefore, we will call $\left | \lambda_{k}^{\mu} \right \rangle$ the pseudo $k$-photon state later.

\subsection{\label{sec:dom}{Dual-optical-mode case}}
Consider two coherent states $\left | \sqrt{\mu} e^{i \phi _{n_{1}}} \right \rangle$ on system $A_1$ and $\left | \sqrt{\mu} e^{i \phi _{n_{2}}} \right \rangle$ on system $A_2$ with $\phi _{n_{1}}$ and $\phi _{n_{2}}$ coming from Eq. (\ref{dp}). Similarly, we employ virtual ancilla qudit systems $\widetilde{A}_1$ and $\widetilde{A}_2$ for each optical mode to record the discrete phase information, and the joint state can be expressed as
\begin{equation}
\begin{aligned}
    \left | \psi_{D}  \right \rangle _{\widetilde{A}_1 \widetilde{A}_2 A_1 A_2} & = \frac{1}{D} \sum_{n_{1} = 0}^{D-1} \sum_{n_{2} = 0}^{D-1}
    \left | n_{1} \right \rangle _{\widetilde{A}_1}
    \left | n_{2} \right \rangle _{\widetilde{A}_2} \\
   &\otimes  \left | \sqrt{\mu} e^{i \phi _{n_{1}}} \right \rangle _{A_1}
    \left | \sqrt{\mu} e^{i \phi _{n_{2}}} \right \rangle _{A_2} \\
& = \frac{1}{D} \sum_{n_{\Delta} = 0}^{D-1} \sum_{n_{2} = 0}^{D-1}
    \left | n_{2} + n_{\Delta} \right \rangle _{\widetilde{A}_1}
    \left | n_{2} \right \rangle _{\widetilde{A}_2} \\
   &\otimes  \left | \sqrt{\mu} e^{i \phi _{n_{2} + n_{\Delta}}} \right \rangle _{A_1}
    \left | \sqrt{\mu} e^{i \phi _{n_{2}}} \right \rangle _{A_2},
\label{dom}
\end{aligned}
\end{equation}
where $n_{\Delta} = n_{1} - n_{2}$, represents the phase difference between the two modes and the symbol + denotes addition modulo $D$.

Considering a fixed value of $n _{\Delta}$, the basis vectors $\big \{ \left | n_{2} + n_{\Delta} \right \rangle _{\widetilde{A}_1} \left | n_{2} \right \rangle _{\widetilde{A}_2} \big \} _{n_{2}=0}^{D-1}$ constitute a subspace in the joint Hilbert space of $\widetilde{A}_1$ and $\widetilde{A}_2$. To obtain the photon number information, we can construct a complementary orthogonal basis $\big \{ \left | n_{\Delta}, k \right \rangle _{\widetilde{A}_1\widetilde{A}_2} \big \}_{k=0}^{D-1}$ for each subspace in a similar way to Eq. (\ref{cb}),
\begin{gather}
    \left | n_{\Delta}, k \right \rangle _{\widetilde{A}_1\widetilde{A}_2} = \frac{1}{\sqrt{D}} \sum_{n_{2}=0}^{D-1} e^{i \frac{2 \pi}{D} n_{2}k}
    \left | n_{2} + n_{\Delta} \right \rangle _{\widetilde{A}_1} \left | n_{2} \right \rangle _{\widetilde{A}_2}, \notag \\
    \left | n_{2} + n_{\Delta} \right \rangle _{\widetilde{A}_1} \left | n_{2} \right \rangle _{\widetilde{A}_2}
  = \frac{1}{\sqrt{D}} \sum_{k=0}^{D-1} e^{-i \frac{2 \pi}{D} n_{2}k} 
    \left | n_{\Delta}, k \right \rangle _{\widetilde{A}_1\widetilde{A}_2}.
\label{dcb}
\end{gather}
This allows us to rewrite Eq. (\ref{dom}) in the form
\begin{equation}
\begin{aligned}
&   \left | \psi_{D}  \right \rangle _{\widetilde{A}_1 \widetilde{A}_2 A_1 A_2}
  = \frac{1}{D\sqrt{D}} \sum_{n_{\Delta}=0}^{D-1} \sum_{n_{2}=0}^{D-1}\sum_{k=0}^{D-1} e^{-i \frac{2 \pi}{D} n_{2}k} \\ 
&   \times  \left | n_{\Delta}, k \right \rangle _{\widetilde{A}_1\widetilde{A}_2}
    \left | \sqrt{\mu} e^{i \phi _{n_{2} + n_{\Delta}}} \right \rangle _{A_1}
    \left | \sqrt{\mu} e^{i \phi _{n_{2}}} \right \rangle _{A_2} \\
& = \sum_{n_{\Delta}=0}^{D-1} \frac{1}{\sqrt{D}} \sum_{k=0}^{D-1} \sqrt[]{P _{k}^{2\mu}} 
    \left | n_{\Delta}, k \right \rangle _{\widetilde{A}_1\widetilde{A}_2}
    \big | n_{\Delta}, \lambda _{k}^{2\mu} \big \rangle _{A_1A_2},
\end{aligned}
\end{equation}
where the normalized state with the overall pseudo photon number $k$ and the relative phase $\frac{2 \pi}{D} n_{\Delta}$ on the system $A_1$ and $A_2$ is given by
\begin{equation}
\begin{aligned}
    \big | n_{\Delta}, \lambda _{k}^{2\mu} \big \rangle _{A_1A_2} & = 
    \frac{1}{D\sqrt[]{P _{k}^{2\mu}}} \hat{U}_{A_1} \big( \frac{2 \pi}{D} n_{\Delta} \big ) \\
&   \circ  \hat{BS} \left ( \sum_{n_{2}=0}^{D-1} e^{-i \frac{2 \pi}{D} n_{2}k}
    \big | \sqrt{2\mu} e^{i \phi _{n_{2}}} \big \rangle_{A_1} 
    \big | 0 \big \rangle _{A_2} \right ) \\
& = \hat{U}_{A_1} \big( \frac{2 \pi}{D} n_{\Delta} \big ) 
    \circ  \hat{BS} \left ( 
    \big | \lambda_{k}^{2\mu} \big \rangle_{A_1} 
    \big | 0 \big \rangle _{A_2} \right ).
\end{aligned}
\end{equation}
To simplify the expression, we use the 50:50 beam splitter $\hat{BS}$ and a phase gate $\hat{U}_{A_1} \left( \phi  \right ) = e^{i \phi a_1^{\dagger} a_1}$, where $a_1$ is the annihilation operator of the mode $A_1$. Additionally, $\circ$ denotes the composition of operators. The state $\big | \lambda_{k}^{2\mu} \big \rangle_{A_1}$ is identical to that in Eq.~(\ref{pks}) except that the intensity parameter $\mu$ is replaced by $2\mu$.

\subsection{\label{sec:bda}{Basis-dependence analysis}}
Under the framework of the original MP-QKD protocol with discrete phase randomization \cite{zeng2022mode}, and according to the decompositions presented in \ref{sec:som} and \ref{sec:dom}, we obtain the encoded logical bases for Alice (Bob's case is analogous) of $Z$ and $\left \{ X_{\theta} \right \}_{\theta = \frac{2 \pi}{D} n, n = 0, 1, 2, ..., \frac{D}{2}-1}$, both with the overall pseudo photon number $k$ as
\begin{equation}
\begin{aligned}
\left | 0 _{k}^{Z} \right \rangle_{A} & = \sum_{n = 0}^{D-1} e^{-i \frac{2 \pi}{D} nk}
\left | 0  \right \rangle _{A_{1}}
\big | \sqrt{\mu} e^{i \frac{2 \pi}{D} n} \big \rangle _{A_{2}}, \\
\left | 1 _{k}^{Z} \right \rangle_{A} & = \sum_{n = 0}^{D-1} e^{-i \frac{2 \pi}{D} nk}
\big | \sqrt{\mu} e^{i \frac{2 \pi}{D} n} \big \rangle _{A_{1}}
\left | 0  \right \rangle _{A_{2}},
\\
\big | 0_{k}^{X_{\theta}} \big \rangle_{A} & = 
\sum_{n = 0}^{D-1} e^{-i \frac{2 \pi}{D} nk}
\big | \sqrt{\mu} e^{i \left ( \frac{2 \pi}{D} n + \theta \right )} \big \rangle _{A_{1}}
\big | \sqrt{\mu} e^{i \frac{2 \pi}{D} n} \big \rangle _{A_{2}}, \\
\big | 1_{k}^{X_{\theta}} \big \rangle_{A} & = 
\sum_{n = 0}^{D-1} e^{-i \frac{2 \pi}{D} nk}
\big | \sqrt{\mu} e^{i \left ( \frac{2 \pi}{D} n + \theta + \pi \right )} \big \rangle _{A_{1}}
\big | \sqrt{\mu} e^{i \frac{2 \pi}{D} n} \big \rangle _{A_{2}}.
\end{aligned}    
\end{equation}
Here, the subscript $A$ denotes Alice, while $A_1$ and $A_2$ represent her subsystems. Therefore, we can write the overall density matrices in the two bases as
\begin{equation}
\begin{aligned}
\rho_{k,k}^{Z} & = \rho_{k,A}^{Z} \otimes \rho_{k,B}^{Z}, \\
\rho_{k,k}^{X_{\theta}} & = \rho_{k,A}^{X_{\theta}} \otimes \rho_{k,B}^{X_{\theta}},
\end{aligned}   
\end{equation}
where the states $\rho_{k,A \left ( B \right )}^{Z}$ and $\rho_{k,A \left ( B \right ) }^{X_{\theta}}$ are given by 
\begin{equation}
\begin{aligned}
\rho_{k,A \left ( B \right )}^{Z} & = \frac{1}{2} \left ( 
\left | 0 _{k}^{Z} \right \rangle \left \langle 0 _{k}^{Z} \right |
+ \left | 1 _{k}^{Z} \right \rangle \left \langle 1 _{k}^{Z} \right |\right )_{A \left ( B \right )},
\\
\rho_{k,A \left ( B \right )}^{X_{\theta}} & = \frac{1}{2} \big ( 
\big | 0_{k}^{X_{\theta}} \big \rangle \big \langle 0_{k}^{X_{\theta}} \big |
+ \big | 1_{k}^{X_{\theta}} \big \rangle \big \langle 1_{k}^{X_{\theta}} \big | \big )_{A \left ( B \right )}.
\end{aligned}
\end{equation}

The security guarantee of the protocol lies in the fact that an eavesdropper Eve cannot distinguish between the two states encoded in the conjugate bases, $Z$ and $X_{\theta}$. When ideal single-photon sources are used, we obtain $\rho_{Z} = \rho_{X_{\theta}}$, in which case the source is said to be \emph{basis-independent}. For further details, see Appendix~\ref{App.A}.

In our case, we can use the fidelity between $\rho_{k,k}^{Z}$ and $\rho_{k,k}^{X_{\theta}}$ to quantify the deviation from the ideal scenario,
\begin{equation}
\begin{aligned}
& F_{k,k}^{\theta} := F \big ( \rho_{k,k}^{Z}, \rho_{k,k}^{X_{\theta}} \big ) = 
tr\sqrt{\sqrt{\rho_{k,k}^{X_{\theta}}} \rho_{k,k}^{Z} \sqrt{\rho_{k,k}^{X_{\theta}}}} \\
& \ge \frac{1}{8} 
\frac{\left| 2 e^{i\theta} S_k(\mu) + S_k(\mu e^{i\theta}) - S_k(-\mu e^{i\theta}) \right|^2}
{\left| S_{k}(\mu) \cdot S_k(2\mu) \right|}, \\
& S_k(\alpha) = \sum_{n=0}^{D-1} e^{i \frac{2\pi}{D} n k} \, 
\exp \left( \alpha \, e^{-i \frac{2\pi}{D}n} \right).
\label{fidelity}
\end{aligned}
\end{equation}
The concrete derivation can be found in Appendix~\ref{App.A}. No positive key rate can be obtained when the fidelity is below $1/\sqrt{2}$ \cite{lo2007security}, which has been confirmed in the BB84 protocol—specifically, the strong basis dependence of multi-photon states renders the system insecure. Since $F_{k,k}^\theta \approx 1$ is achieved only for $k=1$, it can be concluded that in the case of discrete phase randomization, the component contributing to an effective secure key is the pseudo single-photon state. The fidelity of $k=0$ is equal to $1/\sqrt{2}$. However, due to its excessively high error rate, it cannot generate secure keys. Therefore, we will disregard this component in subsequent discussions. 

With different values of $\theta$, we can obtain the overall fidelity between $Z$ and $X$ bases by
\begin{equation}
F_{k,k} = \sum_{\theta} P(\theta) F_{k,k}^\theta,
\label{FXkk}
\end{equation}
where $P(\theta)$ is the conditional probability of choosing the alignment angle $\theta$ in all the sifted $X$-basis data with $\theta^a = \theta^b$, and can be obtained from the experiment directly. For simplicity, we will henceforth assume that $\theta$ follows a uniform distribution, i.e., $P(\theta)=2/D$.

\subsection{\label{sec:level2}{Key rate}}

In the discrete-phase-randomized mode-pairing quantum key distribution, the key rate is given by \cite{zeng2022mode}
\begin{equation}
R = r_{p}\left ( p, l \right ) r_{s} \left \{ q_{1,1}^{Z} 
\left [ 1-H \left ( e_{1,1}^{Z,p} \right ) \right ] - 
f H\left ( E_{\mu, \mu}^Z \right ) \right \},
\label{key rate fomula}
\end{equation}
where $r_p$ denotes the average pairing rate per round, defined as a function of the maximal pairing interval $l$ and the successful click probability $p$ for each pulse, $r_s$ denotes the fraction of $Z$-pairs out of all obtained effective data pairs, $q_{1,1}^{Z}$ is the fraction of $Z$-pairs contributed by pseudo single-photon-pair states $\rho_{1,1}^{Z}$ (i.e., Alice and Bob both emit pseudo single-photon in the paired modes), $e_{1,1}^{Z,p}$ is the corresponding phase error rate, $f$ is the error-correction efficiency, $E_{\mu, \mu}^{Z}$ is the bit error rate of the sifted $Z$-pairs with the signal intensity $\mu$, and $H(x) = -x\log_{2}{(x)} - (1-x)\log_{2}{(1-x)} $ is the binary Shannon entropy function.

In the key rate formula, the parameters $r_p$, $r_s$, and $E_{\mu, \mu}^{Z}$ can be directly measured from experimental data, whereas $q_{1,1}^{Z}$ and $e_{1,1}^{Z,p}$ must be estimated using relevant experimental parameters. In the basis-independent scenario,
e.g., single-photon state source, the phase error rate of the $Z$-basis $e_{1,1}^{Z,p}$ is equal to the bit error rate of the $X$-basis $e_{1,1}^{X,b}$. However, the basis-independence property breaks down for the case of discrete phase randomization. Fortunately, it remains possible to estimate the deviation between $e_{1,1}^{Z,p}$ and $e_{1,1}^{X,b}$ as follows \cite{lo2007security},
\begin{equation}
\begin{aligned}
e_{k,k}^{Z,p} & \le e_{k,k}^{X,b} + 4 \Delta_{k,k} \left ( 1 - \Delta_{k,k} \right ) 
\left ( 1 - 2 e_{k,k}^{X,b} \right ) \\
& + 4 \left ( 1 - 2 \Delta_{k,k} \right ) \sqrt{\Delta_{k,k} \left ( 1 - \Delta_{k,k} \right ) 
e_{k,k}^{X,b} \left ( 1 -  e_{k,k}^{X,b} \right )},
\label{phase bound}
\end{aligned}
\end{equation}
where
\begin{equation}
\Delta_{k,k} = \frac{1 - F_{k,k}}{2Y_{k,k}}.
\label{dev}
\end{equation}
Here, $F_{k,k}$ is given by Eq.~(\ref{FXkk}), and $Y_{k,k}$ is the yield of the state $\rho_{k,k}^X$.

\subsection{\label{sec:level2}{DPR-MP-QKD protocal}}

\begin{figure*}[t] 
\centering
\includegraphics[scale=0.9]{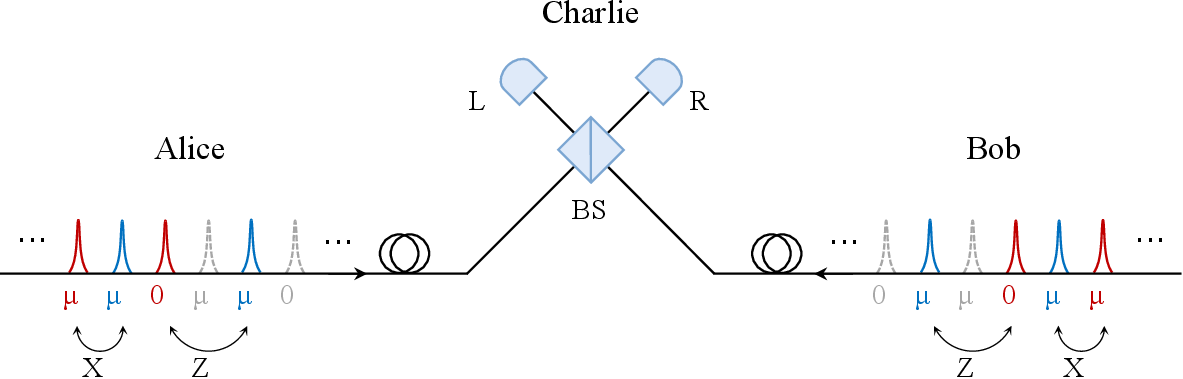}
\caption{Schematic of the discrete-phase-randomized mode-pairing MDI-QKD protocol. Alice and Bob prepare coherent pulses with random intensities from $\left \{ 0, \mu \right \}$ and phases $\phi_i^a$ ($\phi_i^b$) uniformly drawn from $D$ discrete values ${2\pi n/D \mid n = 0, 1, \dots, D-1}$, each chosen with probability $1/D$. Here, $D$ represents the number of discrete phases, which is required to be an even integer. Pulses are sent to Charlie for measurement. Successfully detected pulses (solid lines) within a maximum pairing interval $l$ are paired as leading (red) and trailing (blue) pulses, while pulses that are detected but fail to pair (grey) and pulses with failed detection (dashed lines) are discarded. The basis ($Z$ or $X$) is assigned based on the intensity settings of the paired pulses. For $X$-pairs, alignment angles $\theta^a$ and $\theta^b$ are exchanged, and the data are retained only for $\theta^a = \theta^b$. $Z$-pairs are used for key generation, while the remaining data are reserved for parameter estimation.}
\label{fig:dpr-mp-qkd}
\end{figure*}

The schematic of the DPR-MP-QKD setup is illustrated in Fig.~\ref{fig:dpr-mp-qkd}, with the protocol details elaborated in the following description.

\paragraph{State preparation.}
In the $i$-th location $i\in \left \{ 1, 2, 3, ..., N \right \}$, Alice (Bob) prepares a coherent state
$\big| \sqrt{\mu _{i}^{a}} e^{i \phi _{i}^{a}} \big\rangle$
($\big| \sqrt{\mu _{i}^{b}} e^{i \phi _{i}^{b}} \big\rangle$)
with an intensity $\mu _{i}^{a}$ $(\mu _{i}^{b})$ randomly chosen from $\left \{ 0, \mu \right \}$ and a phase $\phi _{i}^{a}$ $(\phi _{i}^{b})$ uniformly chosen from $D$ different values $\left \{ \frac{2\pi}{D}n \mid n\in 0, 1, 2, ..., D-1 \right \}$ (i.e., each with probability $1/D$). Here, $D$ is the number of discrete phases, and is an even number.

\paragraph{Measurement.}
Alice and Bob send their pulses to an untrusted third party, Charlie, who performs single-photon interference measurements and announces the click patterns for detectors $L$ and $R$. This process is repeated for $N$ rounds, after which Alice and Bob post-process the accumulated data as follows.

\paragraph{Mode pairing.}
For all rounds with a successful detection event (defined as a click in one and only one of the two detectors), Alice and Bob pair two such events provided they occur within the maximal pairing interval $l$. Any pulse that cannot be paired with another within this interval is discarded. The encoded intensities and phases of the two paired rounds then form a data pair.

\paragraph{Basis sifting.}
Based on the intensities of two paired rounds $i$ and $j$, Alice assigns the basis of the data pair as follows: $Z$ for $\mu_i^a + \mu_j^a = \mu$, $X$ for $\mu_i^a = \mu_j^a = \mu$, and ‘0’ for $\mu_i^a = \mu_j^a = 0$. Bob determines the basis using an identical rule. Subsequently, Alice and Bob publicly announce the basis for each pair, and only those pairs for which both announce $Z$ or $X$ are retained. All other data pairs are discarded.

\paragraph{Key mapping.}
For each $Z$-basis pair (or $Z$-pair) at locations $i$ and $j$, Alice sets her raw key bit $\kappa^{a}$ to 0 if $(\mu_i^a, \mu_j^a) = (0, \mu)$ and to 1 if $(\mu_i^a, \mu_j^a) = (\mu, 0)$. In contrast, Bob sets his raw key bit $\kappa^{b}$ to 0 for $(\mu_i^b, \mu_j^b) = (\mu, 0)$ and to 1 for $(\mu_i^b, \mu_j^b) = (0, \mu)$. For each $X$-basis pair ($X$-pair), the raw key is derived from the relative phase. Specifically, Alice computes $\kappa^{a} = \lfloor (\phi_j^a - \phi_i^a) / \pi \rfloor \mod 2$ as her raw key bit, with the alignment angle given by $\theta^{a} = (\phi_j^a - \phi_i^a) \mod \pi$. Bob obtains $\kappa^{b}$ and $\theta^{b}$ in a similar way. Furthermore, Bob adjusts his bit $\kappa^{b}$ based on Charlie's announcement: he keeps $\kappa^{b}$ for click patterns $(L, L)$ or $(R, R)$, and flips it for $(L, R)$ or $(R, L)$. Finally, for all $X$-pairs, Alice and Bob publicly announce their respective alignment angles $\theta^{a}$ and $\theta^{b}$. They retain a pair only if $\theta^{a} = \theta^{b}$; otherwise, the pair is discarded.

\paragraph{Parameter estimation.}
Alice and Bob use the $Z$-pairs to generate a key and the $X$-pairs for parameter estimation. They estimate the fraction of clicked signals $q_{1, 1}^Z$ and the corresponding phase error rate $e_{1, 1}^{Z,p}$ of the $Z$-pairs where Alice and Bob both emit a pseudo single-photon at locations $i$ and $j$, using the data of the $Z$-pairs and $X$-pairs. They also estimate the quantum bit error rate $E_{\mu, \mu}^{Z}$ of the $Z$-pairs.

\paragraph{Key distillation.}
After performing error correction and privacy amplification based on $q_{1, 1}^Z$, $e_{1, 1}^{Z,p}$, and $E_{\mu, \mu}^{Z}$, they ultimately obtained completely identical secret keys.

\section{\label{sec3}Decoy State Estimation}
In this section, we analyze the security of the DPR-MP-QKD protocol. Under discrete phase randomization, the security of the original MP-QKD protocol can no longer be guaranteed. To establish the security of DPR-MP-QKD, we must provide a quantitative estimation method of the proportion of pseudo single-photon state $q_{1,1}^Z$ in the $Z$-basis pairs and the corresponding phase error rate $e_{1, 1}^{Z,p}$ to ensure a positive final key rate can be achieved. To this end, we introduce a discrete phase randomization version of the decoy state method for estimating these parameters. Below, we employ the standard three-intensity decoy state method, which uses a vacuum state, a decoy state with intensity $\nu$ and a signal state with intensity $\mu$, satisfying $0 < \nu < \mu$.

Assume that Alice and Bob have emitted $N$ pulses, where $N \to \infty$. At each position $i$, Alice and Bob each independently and randomly select one of three possible intensities with probabilities $s_{0}, s_{\nu}$ and $s_{\mu}$, respectively, where $s_{0} + s_{\nu} + s_{\mu} = 1$. As in original MP-QKD, a specific fixed pairing strategy $\vec{\chi}$ defined over all locations is considered here, which also includes the locations with unsuccessful clicks. Alice and Bob pair two locations, $i$ and $j$, according to the pairing setting $\vec{\chi}$ as a single round of QKD. The intensity vector of the $\left ( i, j \right ) $ pair is denoted as

\begin{equation}
\vec{\mu} = \left ( \mu_{i}^{a} + \mu_{j}^{a}, \mu_{i}^{b} + \mu_{j}^{b} \right ),
\end{equation}
where $\mu_{i}^{a\left ( b \right ) }, \mu_{j}^{a\left ( b \right ) } \in \left \{ 0, \nu, \mu \right \}$ and, therefore, $\mu_{i}^{a\left ( b \right ) } + \mu_{j}^{a\left ( b \right ) } \in \left \{ 0, \nu, \mu, 2\nu, \nu + \mu, 2\mu \right \}$. The probability of Alice and Bob sending out intensities $\vec{\mu}$ for the $(i,j)$ pair is as follows,
\begin{equation}
q^{\vec{\mu}} = \sum_{\big ( \mu_{i}^{a} + \mu_{j}^{a}, \mu_{i}^{b} + \mu_{j}^{b} \big )
= \vec{\mu}} s_{\mu_{i}^{a}} s_{\mu_{j}^{a}} s_{\mu_{i}^{b}} s_{\mu_{j}^{b}}.
\end{equation}

Referring to section \ref{sec:dom}, we can see that Alice and Bob can perform an overall pseudo photon number measurement on ancilla system $A_1 A_2$ for each pair of locations $(i,j)$. We denote the measurement result of the pseudo photon number as $\vec{k} = (k^a, k^b)$. For a given intensity $\vec{\mu}$, the joint probability of Alice and Bob obtaining the pseudo photon number measurement outcome $\vec{k}$ is 
\begin{equation}
\begin{aligned}
Pr(\vec{k} \mid \vec{\mu} ) & = 
e^{-(\mu_{i}^{a} + \mu_{j}^{a})} \sum_{m_1 = 0}^{\infty} 
\frac{(\mu_{i}^{a} + \mu_{j}^{a}) ^{\left ( m_1D+k^a \right ) }} {\left ( m_1D+k^a \right ) ! } \\
& \times  
e^{-(\mu_{i}^{b} + \mu_{j}^{b})} \sum_{m_2 = 0}^{\infty} 
\frac{(\mu_{i}^{b} + \mu_{j}^{b}) ^{\left ( m_2D+k^b \right ) }} {\left ( m_2D+k^b \right ) ! },
\end{aligned}
\end{equation}
which is the product of two factors due to the independence of Alice's and Bob's intensity settings. As $D$ approaches infinity, each factor converges to a Poisson distribution. 

After mode pairing, for a pulse pair $(i,j)$, Alice and Bob assign the bases based on the intensity values they selected. If the intensity setting of this pair $\vec{\mu}$ meets the requirement,
\begin{equation}
\mu_{i}^{a} \mu_{j}^{a} = \mu_{i}^{b} \mu_{j}^{b} = 0, \,
\mu_{i}^{a} + \mu_{j}^{a} + \mu_{i}^{b} + \mu_{j}^{b}\ne 0,
\end{equation}
then it belongs to the $Z$-basis. Here, our analysis mainly focuses on decoy state estimation in the $Z$-basis, and the results for the $X$-basis can be readily derived in a similar way.

After basis sifting, assume that $M$ rounds of sifted $Z$-basis pairs are shared between Alice and Bob, among which $T$ rounds are error-containing. These can be further categorized according to intensity settings and pseudo photon numbers: let $M^{\vec{\mu}}$ denote the number of rounds detected with intensity setting $\vec{\mu}$, among which $T^{\vec{\mu}}$ rounds are erroneous, let $M_{\vec{k}}$ represent the total number of rounds detected under pseudo $\vec{k}$-photon number, with $T^{\vec{\mu}}$ being the corresponding erroneous rounds, and let $M_{\vec{k}}^{\vec{\mu}}$ and $T_{\vec{k}}^{\vec{\mu}}$ denote the number of rounds and erroneous rounds, respectively, that satisfy both conditions. Based on this, we obtain the following relations,
\begin{equation}
\begin{aligned}
M & = \sum_{\vec{\mu}} M^{\vec{\mu}} = \sum_{\vec{k}} M_{\vec{k}} =
\sum_{\vec{\mu}} \sum_{\vec{k}} M_{\vec{k}}^{\vec{\mu}}, \\
T & = \sum_{\vec{\mu}} T^{\vec{\mu}} = \sum_{\vec{k}} T_{\vec{k}} =
\sum_{\vec{\mu}} \sum_{\vec{k}} T_{\vec{k}}^{\vec{\mu}}.
\end{aligned}    
\end{equation}

Since a total of $N$ pulses have been emitted, the expected total paired number
can be obtained from $N_p = \left \lfloor N/2 \right \rfloor$. The expected number of rounds with intensity setting $\vec{\mu}$ is given by $N^{\vec{\mu}} = q^{\vec{\mu}} N_p$ and for a given intensity setting $\vec{\mu}$, the expected number of rounds containing a pseudo photon number $\vec{k}$  can be calculated by $N_{\vec{k}}^{\vec{\mu}} = Pr(\vec{k} \mid \vec{\mu}) N^{\vec{\mu}}$. Thus, we can define the ‘gain’ and ‘yield’ as
\begin{equation}
\begin{aligned}
Q^{\vec{\mu}} & = \frac{M^{\vec{\mu}}}{N^{\vec{\mu}}}, \,\, 
Q^{\vec{\mu}} \cdot E^{\vec{\mu}} = \frac{T^{\vec{\mu}}}{N^{\vec{\mu}}}, \,\, \\
Y_{\vec{k}}^{\vec{\mu}} & = \frac{M_{\vec{k}}^{\vec{\mu}}}{N_{\vec{k}}^{\vec{\mu}}}, \,\, 
Y_{\vec{k}}^{\vec{\mu}} \cdot e_{\vec{k}}^{\vec{\mu}} = \frac{T_{\vec{k}}^{\vec{\mu}}}{N_{\vec{k}}^{\vec{\mu}}}, \,\, 
\end{aligned}
\end{equation}
where $E^{\vec{\mu}}$ and $e_{\vec{k}}^{\vec{\mu}}$ are the corresponding quantum bit error rate (QBER). In standard MP-QKD, there exists a fundamental assumption that
\begin{equation}
\begin{aligned}
Y_{\vec{k}}^{\vec{\mu_1}} & = Y_{\vec{k}}^{\vec{\mu_2}}, \\
e_{\vec{k}}^{\vec{\mu_1}} & = e_{\vec{k}}^{\vec{\mu_2}},
\end{aligned}
\end{equation}
where $\vec{\mu}=(\mu^a=\mu_i^a+\mu_j^a, \mu^b=\mu_i^b+\mu_j^b)$ and $\vec{k}=(k^a, k^b)$. Nevertheless, this assumption is no longer valid in the regime of discrete phase randomization as
\begin{equation}
\left | \lambda_{\vec{k}}^{\vec{\mu_1}} \right \rangle \ne \left | \lambda_{\vec{k}}^{\vec{\mu_2}} \right \rangle,
\end{equation}
where $\big | \lambda_{\vec{k}}^{\vec{\mu}} \big \rangle$ represents the joint state of Alice and Bob, corresponding to Alice using intensity $\mu^a$ with pseudo $k^a$-photon state and Bob using intensity $\mu^b$ with pseudo $k^b$-photon state. However, it is possible to bound the differences in yield and error rates between different intensity settings as
\begin{equation}
\begin{aligned}
\left | Y_{k^a, k^b}^{\alpha, \mu} - Y_{k^a, k^b}^{\alpha, \nu} \right | 
& \le \sqrt{1-F_{\mu\nu}^2},\\
\left | Y_{k^a, k^b}^{\alpha, \mu} e_{k^a, k^b}^{\alpha, \mu} - 
Y_{k^a, k^b}^{\alpha, \nu} e_{k^a, k^b}^{\alpha, \nu} \right | 
& \le \sqrt{1-F_{\mu\nu}^2},
\end{aligned}
\end{equation}
where $\alpha$ represents one of the possible intensity values. For subsequent convenience, we expand the vectors of intensity and pseudo photon number. $F_{\mu\nu}$ is given as follows, 
\begin{equation}
F_{\mu\nu} = \frac{\sum_{m=0}^{\infty} \frac{\left ( \mu\nu \right ) ^{mD/2} }{\left ( mD \right )!} }{\sqrt{
\sum_{m_1=0}^{\infty}\frac{\mu^{m_1D} }{\left ( m_1D \right )!}
\sum_{m_2=0}^{\infty}\frac{\nu^{m_2D} }{\left ( m_2D \right )!}}}.
\end{equation}
The derivation of these bounds can be found in Appendix~\ref{App.B}. In discrete-phase-randomized MP-QKD, we need to estimate the yield $Y_{k^a, k^b}^{\mu^a, \mu^b}$ and the error rate $e_{k^a, k^b}^{\mu^a, \mu^b}$. The relations between gain and yield are given by
\begin{equation}
\begin{aligned}
Q^{\mu^a,\mu^b} & = \sum_{k^a,k^b=0}^{D-1} P_{k^a}^{\mu^a} P_{k^b}^{\mu^b} 
Y_{k^a,k^b}^{\mu^a,\mu^b}, \\
Q^{\mu^a,\mu^b} E^{\mu^a,\mu^b} & = \sum_{k^a,k^b=0}^{D-1} P_{k^a}^{\mu^a} P_{k^b}^{\mu^b} 
Y_{k^a,k^b}^{\mu^a,\mu^b} e_{k^a,k^b}^{\mu^a,\mu^b},
\label{dpr-qkd}
\end{aligned}
\end{equation}
where $P_{k}^{\mu}$ is given in Eq.~(\ref{dpr-pro}). The gain $Q^{\mu^a,\mu^b}$ and QBER $E^{\mu^a,\mu^b}$ can be obtained directly in experiment. Note that the correctness of Eq.~(\ref{dpr-qkd}) remains unaffected by the specific value of $N_p$. If we rescale $N_p$ as $N_{p}' = c N_p$, the corresponding gain and yield become $Q^{\mu^a,\mu^b}/c$ and $Y_{k^a,k^b}^{\mu^a,\mu^b}/c$, respectively.

The estimation of the yield $Y_{k^a,k^b}^{\mu^a,\mu^b}$ and the error rate $e_{k^a,k^b}^{\mu^a,\mu^b}$ is similar to the conventional MDI-QKD. Here we only introduce the estimation of the yield $Y_{k^a,k^b}^{\mu^a,\mu^b}$, and the estimation of $e_{k^a,k^b}^{\mu^a,\mu^b}$ is similar. Note that the first expression in Eq.~(\ref{dpr-qkd}) can be reformulated as
\begin{equation}
Q^{\mu^a,\mu^b} = \sum_{k^a=0}^{D-1} P_{k^a}^{\mu^a} Y_{k^a}^{\mu^a,\mu^b},
\end{equation}
where
\begin{equation}
Y_{k^a}^{\mu^a,\mu^b} = \sum_{k^b=0}^{D-1} P_{k^b}^{\mu^b} Y_{k^a,k^b}^{\mu^a,\mu^b}.
\label{dpr-yie}
\end{equation}

For simplicity, we define
\begin{equation}
\epsilon = \sqrt{1-F_{\mu\nu}^2}.
\end{equation}

From Eq.~(\ref{dpr-yie}), we can obtain the following,
\begin{equation}
\begin{aligned}
\left | Y_{k^a}^{\mu_1^a,\mu^b} - Y_{k^a}^{\mu_2^a,\mu^b} \right | & = \left |
\sum_{k^b = 0}^{D-1} P_{k^b}^{\mu^b} \left (  
Y_{k^a,k^b}^{\mu_1^a,\mu^b} - Y_{k^a,k^b}^{\mu_2^a,\mu^b} \right ) \right |   \\
& \le \sum_{k^b = 0}^{D-1} P_{k^b}^{\mu^b}
\left | Y_{k^a,k^b}^{\mu_1^a,\mu^b} - Y_{k^a,k^b}^{\mu_2^a,\mu^b} \right | \\
& \le \sum_{k^b = 0}^{D-1} P_{k^b}^{\mu^b} \epsilon = \epsilon,
\end{aligned}
\end{equation}
where the last inequality is guaranteed by the property $\sum_{k = 0}^{D-1} P_{k}^{\mu} = 1$. Based on the preceding analysis, the upper and lower bounds of $Y_{k^a}^{\mu, \mu_b}$ can be estimated subject to the following constraints,
\begin{equation}
\begin{aligned}
Q^{\mu,\mu^b} & = \sum_{k^a = 0}^{D-1} P_{k^a}^{\mu} Y_{k^a}^{\mu,\mu^b}, \\
Q^{\mu^a,\mu^b} & = \sum_{k^a=0}^{D-1} P_{k^a}^{\mu^a} Y_{k^a}^{\mu^a,\mu^b} = 
\sum_{k^a=0}^{D-1} P_{k^a}^{\mu^a} Y_{k^a}^{\mu,\mu^b} \pm \epsilon,\\
0 & \le Y_{k^a}^{\mu, \mu_b} \le 1.
\end{aligned}    
\end{equation}

Upon estimating the range of $ Y_{k^a}^{\mu, \mu_b}$ for all $\mu^b$, the upper and lower bounds of $Y_{k^a, k^b}^{\mu, \mu}$ can be derived under the following constraints,
\begin{equation}
\begin{aligned}
Y_{k^a}^{\mu,\mu} & = \sum_{k^b = 0}^{D-1} P_{k^b}^{\mu} Y_{k^a, k^b}^{\mu,\mu}, \\
Y_{k^a}^{\mu,\mu^b} & = \sum_{k^b=0}^{D-1} P_{k^b}^{\mu^b} Y_{k^a, k^b}^{\mu,\mu^b} = 
\sum_{k^b=0}^{D-1} P_{k^b}^{\mu^b} Y_{k^a, k^b}^{\mu,\mu} \pm \epsilon,\\
0 & \le Y_{k^a, k^b}^{\mu, \mu} \le 1.
\end{aligned}    
\end{equation}
Subsequently, we can derive the proportion of pseudo single-photon-pair state in the Z-basis using the following formula,
\begin{equation}
q_{1,1}^Z = \frac{(P_{1}^{\mu})^2 Y_{1, 1}^{\mu,\mu}}{Q^{\mu, \mu}},
\end{equation}
where the numerator represents the gain of pseudo single-photon pairs, and the denominator represents the gain of signal pairs.

Having obtained the estimate for the bit error rate $e_{1, 1}^{X,b}$ in the X-basis using a similar approach, we can utilize Eq.~(\ref{phase bound}) to bound the phase error rate $e_{1, 1}^{Z,p}$ in the Z-basis.

\section{\label{sec4}Simulation result}
In this section, we simulate the performance of DPR-MP-QKD as a function of the communication distance under different pairing intervals $l$ and different numbers of discrete phases $D$. The intensity $\mu$ is varied within the range $[0, 0.5]$ to optimize the key rate. The simulation parameters are summarized in Table~\ref{tab:para}.

\begin{table}[b]
\caption{\label{tab:para}%
Simulation Parameters}
\begin{ruledtabular}
\begin{tabular}{ccccc}
Parameter & Symbol & Value \\
\colrule
Dark count rate & $p_d$ & $1.2 \times 10^{-8}$ \\
Error correction efficiency & $f$ & 1.15 \\
Detector efficiency & $\eta_d$ & 0.2 \\
Misalignment error rate & $e_d$ & 0.04 \\
Fiber loss & $\alpha$ & 0.2 dB/km \\
\end{tabular}
\end{ruledtabular}
\end{table}

\begin{figure}[h!]
\includegraphics[scale=0.5]{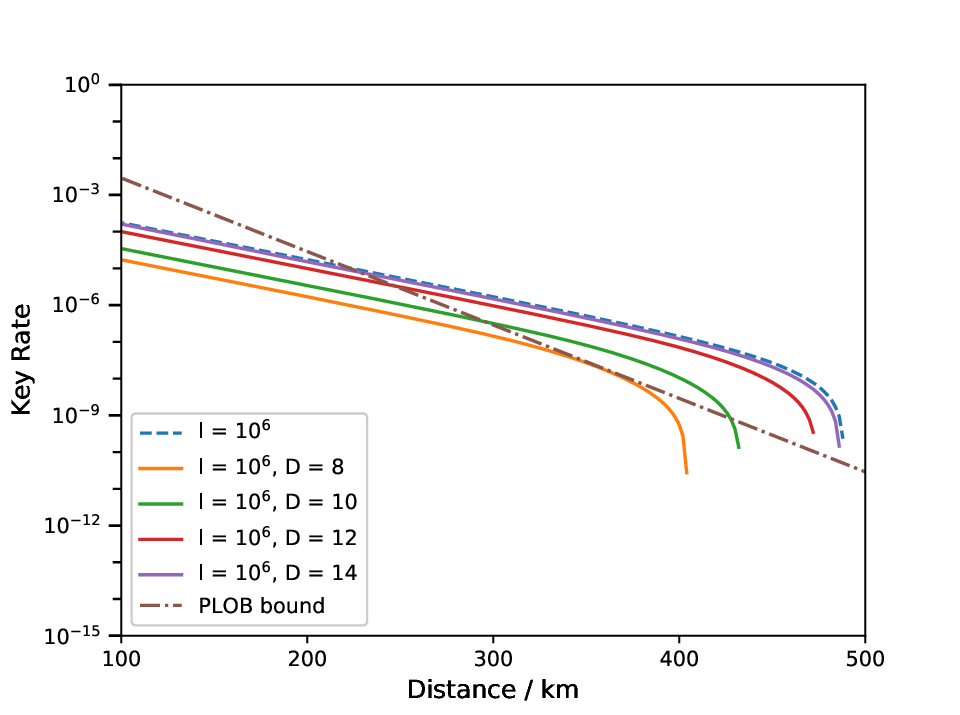}
\caption{\label{fig1} The relation between the key rate and the transmission distance for continuously random phases and discrete phases with maximal pairing interval $l=10^6$. The dashed line is the key rate for continuously random phases and the solid lines from left to right are for 8, 10, 12, and 14 discrete phases, respectively. The dot-dashed line is the PLOB bound.}
\end{figure}

\begin{figure}
\includegraphics[scale=0.5]{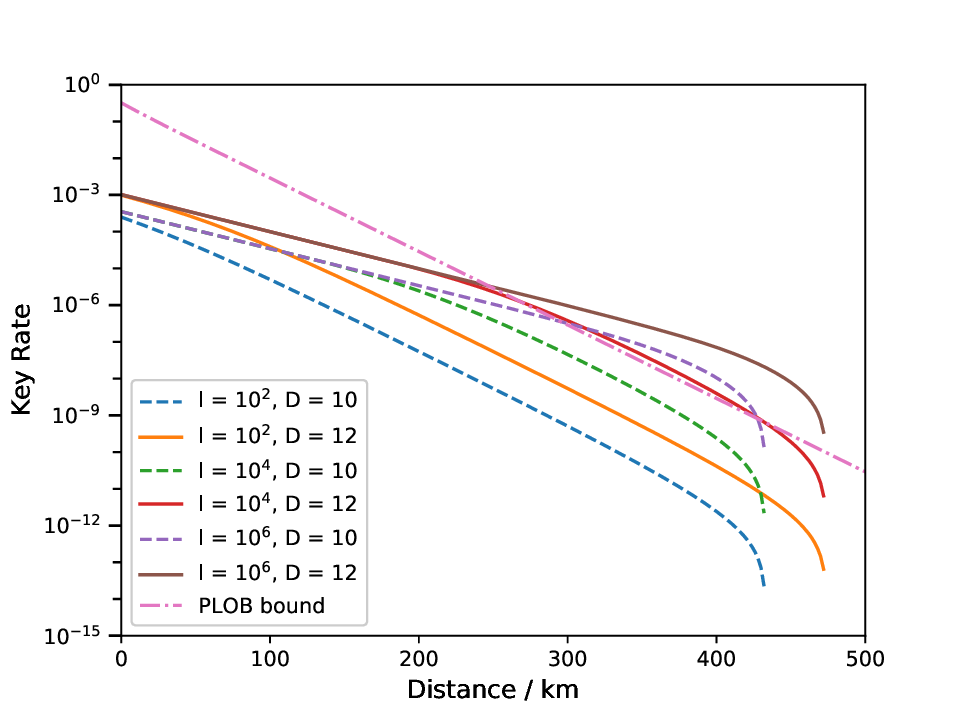}
\caption{\label{fig2}  The relation between the key rate and the transmission distance for discrete phase with maximal pairing interval $l=10^2, 10^4$ and $10^6$. The dashed line is the key rate with discrete phase number $D=10$ and the solid lines is the key rate with discrete phase number $D=12$. The dot-dashed line is the PLOB bound.}
\end{figure}

In Fig.~\ref{fig1}, the key rate performance under the maximum pairing interval of $l=10^6$ is plotted as a function of transmission distance for both continuous phase randomization and discrete randomization with various phase numbers (8, 10, 12, and 14 phases). The results demonstrate that as the number of discrete phases increases, the key rate performance under discrete phase randomization progressively approaches that of continuous phase randomization, with 14 phases already achieving close approximation. Moreover, under this configuration, nearly all discrete phase cases demonstrate the capability to surpass the bound, highlighting the superior performance of DPR-MP-QKD. In Fig.~\ref{fig2}, we present the key rate performance under different maximum pairing intervals $l$ (set to $10^2$, $10^4$ and $10^6$), with a fixed discrete phase number $D=10$ and $12$. We observe that although the key rate improves with larger $l$ and demonstrates the capability to surpass the PLOB bound, aligning with findings in the original MP-QKD \cite{zeng2022mode}, the maximum achievable communication distance remains limited by the discrete phase number. More details on the simulation model are provided in Appendix~\ref{App.C}.

\section{\label{sec5}conclusion}
In this work, to address the issue that the continuous phase randomization assumption in the original MP-QKD protocol is experimentally infeasible, thus compromising its security, we propose a discrete-phase-randomized MP-QKD protocol as a solution. By analyzing the basis-dependence of the DPR-MP-QKD source, we demonstrate that only pseudo single-photon states contribute positively to the key rate. Security analysis via the decoy state method confirms that discrete phase randomization does not compromise the protocol's security. Numerical simulations show that as the number of discrete phases $D$ increases, the key rate progressively approaches that of the continuous case, nearly reaching convergence at $D=14$. Compared to conventional continuous phase randomization, which consumes an infinite number of random bits, our scheme significantly reduces this requirement to as few as 4 bits, which are sufficient to encode the 14 discrete phase values since $2^4 = 16 > 14$. Furthermore, while the maximum secure communication distance is limited by the value of $D$, the key rate retains the capability to surpass the PLOB bound as the maximum pairing interval $l$ increases, a characteristic consistent with the original MP-QKD protocol \cite{zeng2022mode}. This demonstrates the superior performance of our DPR-MP-QKD scheme.

In future work, we may extend our analysis to encompass additional source imperfections in mode-pairing quantum key distribution. One potential direction involves considering incorrectly prepared discrete phases of the form $\big | \alpha e^{2 \pi n/D \pm \delta }  \big \rangle $, where $\delta$ represents a small deviation from the ideal discrete phase values. The fidelity calculation can be adapted accordingly to incorporate such phase inaccuracies. Beyond the asymptotic analysis presented here, extending our work to the finite-key regime is essential for practical deployment, as it accounts for statistical fluctuations due to finite data blocks and provides a realistic security guarantee \cite{wang2023tight}. Another potential direction involves extending our work to the scenario with asymmetric channels \cite{lu2024asymmetric, wang2025asymmetric, li2025asymmetric}.

\begin{acknowledgments}
This work was supported by the Natural Science Foundation of Shanghai under Grant 25ZR1402098, Shanghai Science and Technology Project
under Grant 24LZ1401600, the National Natural Science Foundation of China under Grant 62373155, and the startup fund from East China University of Science and Technology under Grant YH0142234.
\end{acknowledgments}

\appendix  

\section{Basis Dependence} \label{App.A}
We first consider the case where Alice and Bob, each holding a single-photon source, encode information in the single-photon subspace of two orthogonal optical modes $A_{1}\left ( B_{1} \right )$ and $A_{2}\left ( B_{2} \right )$, thereby forming a dual-rail qubit. This encoding naturally leads to the definition of the $Z$-basis as
\begin{equation}
\begin{aligned}
\left | Z\left ( 0 \right )   \right \rangle & = \left | 0  \right \rangle_{A_{1}} 
\left | 1 \right \rangle_{A_{2}},\\
\left | Z\left ( 1 \right )   \right \rangle & = \left | 1  \right \rangle_{A_{1}} 
\left | 0 \right \rangle_{A_{2}}.
\end{aligned}
\end{equation}
The $\left \{ X_{\theta} \right \}_{\theta \in \left [ 0, \pi \right )}$ basis is defined as
\begin{equation}
\begin{aligned}
\left | X_{\theta} \left ( + \right ) \right \rangle & = \frac{1}{\sqrt{2}} \left [
\left | Z \left ( 0 \right ) \right \rangle + e^{i \theta}
\left | Z \left ( 1 \right ) \right \rangle \right ]_{A_{1}A_{2}} \\
& = \frac{1}{\sqrt{2}} \left [
\left | 0  \right \rangle \left | 1 \right \rangle + e^{i \theta}
\left | 1  \right \rangle \left | 0 \right \rangle \right ]_{A_{1}A_{2}},
\\
\left | X_{\theta} \left ( - \right ) \right \rangle & = \frac{1}{\sqrt{2}} \left [
\left | Z \left ( 0 \right ) \right \rangle - e^{i \theta}
\left | Z \left ( 1 \right ) \right \rangle \right ]_{A_{1}A_{2}} \\
& = \frac{1}{\sqrt{2}} \left [
\left | 0  \right \rangle \left | 1 \right \rangle - e^{i \theta}
\left | 1  \right \rangle \left | 0 \right \rangle \right ]_{A_{1}A_{2}}.
\end{aligned}
\end{equation}
Thus,
\begin{equation}
\rho_Z = \rho_{X_\theta } = \frac{1}{2} \left ( 
\left | 0  \right \rangle \left | 1  \right \rangle 
\left \langle 0 \right | \left \langle 1 \right | +
\left | 1  \right \rangle \left | 0  \right \rangle 
\left \langle 1 \right | \left \langle 0 \right |
\right )_{A_1A_2},
\end{equation}
where we can see that the fidelity between the $Z$ and $X_\theta$ bases is 1. Therefore, the single-photon source is a \emph{basis-independent} source. Here we primarily focus on Alice's side, as Bob's case follows analogously.

We now consider the discrete-phase-randomized mode-pairing protocol proposed in this work. For clarity, we reiterate the definitions of the normalized logically encoded $Z$ and $\left \{ X_{\theta} \right \}_{\theta = \frac{\pi}{D} n, n = 0, 1, 2, ..., D-1}$ bases, both with the overall pseudo photon number $k$ as follows,
\begin{equation}
\begin{aligned}
\left | 0 _{k}^{Z} \right \rangle & = \frac{\sum_{n = 0}^{D-1} e^{-i \frac{2 \pi}{D} nk}
\left | 0 \right \rangle
\big | \sqrt{\mu} e^{i \frac{2 \pi}{D} n} \big \rangle}
{\sqrt{D e^{-\mu} \sum_{n=0}^{D-1} e^{-i \frac{2 \pi}{D} nk} e^{\mu e^{i \frac{2 \pi}{D} n}}}}, 
\\
\left | 1 _{k}^{Z} \right \rangle & = \frac{\sum_{n = 0}^{D-1} e^{-i \frac{2 \pi}{D} nk}
\big | \sqrt{\mu} e^{i \frac{2 \pi}{D} n} \big \rangle
\left | 0 \right \rangle}
{\sqrt{D e^{-\mu} \sum_{n=0}^{D-1} e^{-i \frac{2 \pi}{D} nk} e^{\mu e^{i \frac{2 \pi}{D} n}}}},
\\
\big | 0_{k}^{X_{\theta}} \big \rangle & = \frac{\sum_{n = 0}^{D-1} e^{-i \frac{2 \pi}{D} nk}
\big | \sqrt{\mu} e^{i \left ( \frac{2 \pi}{D} n + \theta \right )} \big \rangle
\big | \sqrt{\mu} e^{i \frac{2 \pi}{D} n} \big \rangle}
{\sqrt{D e^{-2\mu} \sum_{n=0}^{D-1} e^{i \frac{2 \pi}{D} nk} e^{2\mu e^{-i \frac{2 \pi}{D} n}}}},
\\
\big | 1_{k}^{X_{\theta}} \big \rangle & = \frac{\sum_{n = 0}^{D-1} e^{-i \frac{2 \pi}{D} nk}
\big | \sqrt{\mu} e^{i \left ( \frac{2 \pi}{D} n + \theta + \pi \right )} \big \rangle
\big | \sqrt{\mu} e^{i \frac{2 \pi}{D} n} \big \rangle}
{\sqrt{D e^{-2\mu} \sum_{n=0}^{D-1} e^{i \frac{2 \pi}{D} nk} e^{2\mu e^{-i \frac{2 \pi}{D} n}}}},
\end{aligned}
\end{equation}
where the denominators are the normalization factors.

To assess the fidelity between the two states in the two bases, we compute the relevant inner products among these four quantum states,
\begin{equation}
\begin{aligned} 
& \big \langle 0 _{k}^{Z} \big | 0_{k}^{X_{\theta}} \big \rangle = 
\big \langle 0 _{k}^{Z} \big | 1_{k}^{X_{\theta}} \big \rangle = 
\frac{S_k(\mu)}{\sqrt{S_{k}(\mu) \cdot S_k(2\mu)}},
\\ 
& \big \langle 1 _{k}^{Z} \big | 0_{k}^{X_{\theta}} \big \rangle =
\frac{S_k(\mu e^{i\theta})}{\sqrt{S_{k}(\mu) \cdot S_k(2\mu)}},
\\
& \big \langle 1 _{k}^{Z} \big | 0_{k}^{X_{\theta}} \big \rangle =
\frac{S_k(-\mu e^{i\theta})}{\sqrt{S_{k}(\mu) \cdot S_k(2\mu)}},
\\
& S_k(\alpha) = \sum_{n=0}^{D-1} e^{i \frac{2\pi}{D} n k} \, 
\exp \left( \alpha \, e^{-i \frac{2\pi}{D}n} \right).
\end{aligned}
\end{equation}
The detailed calculation method for the normalization factors and the inner products can refer to Ref.~\cite{cao2015discrete}. 

The specific fidelity calculation formula is given as follows,
\begin{equation}
\scalebox{0.85}{$\displaystyle
\begin{aligned}
F_{k,k}^{\theta} & = F \big ( \rho_{k,k}^{Z}, \rho_{k,k}^{X_{\theta}} \big ) \\
& = F \big ( \big ( 
\left | 0 _{k}^{Z} \right \rangle \left \langle 0 _{k}^{Z} \right | +
\left | 1 _{k}^{Z} \right \rangle \left \langle 1 _{k}^{Z} \right |
\big )_A
\otimes 
\big ( 
\left | 0 _{k}^{Z} \right \rangle \left \langle 0 _{k}^{Z} \right | +
\left | 1 _{k}^{Z} \right \rangle \left \langle 1 _{k}^{Z} \right |
\big )_B, \\
& \big ( 
\big | 0_{k}^{X_{\theta}} \big \rangle \big \langle 0_{k}^{X_{\theta}} \big | +
\big | 1_{k}^{X_{\theta}} \big \rangle \big \langle 1_{k}^{X_{\theta}} \big |
\big )_A \otimes 
\big ( 
\big | 0_{k}^{X_{\theta}} \big \rangle \big \langle 0_{k}^{X_{\theta}} \big | +
\big | 1_{k}^{X_{\theta}} \big \rangle \big \langle 1_{k}^{X_{\theta}} \big |
\big )_B
\big ) \\
& =  F \big ( 
\left | 0 _{k}^{Z} \right \rangle \left \langle 0 _{k}^{Z} \right | +
\left | 1 _{k}^{Z} \right \rangle \left \langle 1 _{k}^{Z} \right |,
\big | 0_{k}^{X_{\theta}} \big \rangle \big \langle 0_{k}^{X_{\theta}} \big | +
\big | 1_{k}^{X_{\theta}} \big \rangle \big \langle 1_{k}^{X_{\theta}} \big |
\big )^2 \\
& \ge F \big ( 
\left | 0 \right \rangle \left | 0 _{k}^{Z} \right \rangle + e^{2i \theta}
\left | 1 \right \rangle \left | 1 _{k}^{Z} \right \rangle,
\left | +_\theta  \right \rangle \big | 0_{k}^{X_{\theta}} \big \rangle + 
\left | -_\theta  \right \rangle \big | 1_{k}^{X_{\theta}} \big \rangle
\big )^2 \\
& = \frac{1}{4} \left | \left (
\left \langle 0 \right | \left \langle 0 _{k}^{Z}\right | + e^{-2i \theta}
\left \langle 1 \right | \left \langle 1 _{k}^{Z}\right |
\right ) \big (
\left | +_\theta  \right \rangle \big | 0_{k}^{X_{\theta}} \big \rangle + 
\left | -_\theta  \right \rangle \big | 1_{k}^{X_{\theta}} \big \rangle
\big ) \right |^2 \\
& = \frac{1}{8} \left | 
2 e^{i \theta} \big \langle 0 _{k}^{Z} \big | 0_{k}^{X_{\theta}} \big \rangle + 
\big \langle 1 _{k}^{Z} \big | 0_{k}^{X_{\theta}} \big \rangle -
\big \langle 1 _{k}^{Z} \big | 1_{k}^{X_{\theta}} \big \rangle
\right |^2 \\
& = \frac{1}{8} 
\frac{\left| 2 e^{i\theta} S_k(\mu) + S_k(\mu e^{i\theta}) - S_k(-\mu e^{i\theta}) \right|^2}
{\left| S_{k}(\mu) \cdot S_k(2\mu) \right|},
\end{aligned}
$}
\end{equation}
where the states $\left | 0 \right \rangle, \left | 1 \right \rangle$ and $\left | \pm _\theta \right \rangle = \left ( \left | 0  \right \rangle \pm e^{i \theta} \left | 1  \right \rangle \right ) / \sqrt{2}$ are the eigenstates of the $Z$ and $X_\theta$ bases, respectively. The inequality follows from the maximal fidelity property of mixed states, which states that their fidelity equals the maximum fidelity achievable over all possible purifications. This can be intuitively understood by noting that the following two maximally entangled states are identical,
\begin{equation}
\left | 0 \right \rangle \left | 0 \right \rangle + e^{2i \theta}
\left | 1 \right \rangle \left | 1 \right \rangle = 
\left | +_\theta \right \rangle \left | +_\theta \right \rangle +
\left | -_\theta \right \rangle \left | -_\theta \right \rangle.
\end{equation}

Thus, Eq.~(\ref{fidelity}) in the main text is proved.

\section{Decoy-State Parameter Deviation} \label{App.B}
In this section, we provide a detailed analysis of the deviations in decoy state yields and error rates. Similar to the approach in the preceding section, our derivation incorporates several key results from Ref.~\cite{cao2015discrete}. 

Based on the quantum coin idea \cite{gottesman2004security}, we can obtain
\begin{equation}
\begin{gathered}
\sqrt{Y_{k^a, k^b}^{\alpha, \mu} \, Y_{k^a, k^b}^{\alpha, \nu}}  + 
\sqrt{(1 - Y_{k^a, k^b}^{\alpha, \mu}) \, (1 - Y_{k^a, k^b}^{\alpha, \nu})} \\
\ge F\left ( \left | \lambda_{k^a}^{\alpha}  \right \rangle
\left | \lambda_{k^b}^{\mu}  \right \rangle,
\left | \lambda_{k^a}^{\alpha}  \right \rangle
\left | \lambda_{k^b}^{\nu}  \right \rangle \right ),
\\
\sqrt{Y_{k^a, k^b}^{\alpha, \mu} e_{k^a, k^b}^{\alpha, \mu} \, 
Y_{k^a, k^b}^{\alpha, \nu} e_{k^a, k^b}^{\alpha, \nu}}  + 
\\
\sqrt{(1 - Y_{k^a, k^b}^{\alpha, \mu} e_{k^a, k^b}^{\alpha, \mu}) \, 
(1 - Y_{k^a, k^b}^{\alpha, \nu} e_{k^a, k^b}^{\alpha, \nu})} \\
\ge F\left ( \left | \lambda_{k^a}^{\alpha}  \right \rangle
\left | \lambda_{k^b}^{\mu}  \right \rangle,
\left | \lambda_{k^a}^{\alpha}  \right \rangle
\left | \lambda_{k^b}^{\nu}  \right \rangle \right ).
\end{gathered}
\end{equation}

The right-hand side can be simplified as
\begin{equation}
F\left ( \left | \lambda_{k^a}^{\alpha}  \right \rangle
\left | \lambda_{k^b}^{\mu}  \right \rangle,
\left | \lambda_{k^a}^{\alpha}  \right \rangle
\left | \lambda_{k^b}^{\nu}  \right \rangle \right )
= F\left ( \left | \lambda_{k^b}^{\mu}  \right \rangle,
\left | \lambda_{k^b}^{\nu}  \right \rangle \right ) 
\ge F_{\mu \nu}.
\end{equation}
The first inequality holds since the first subsystems of both states are identical, while the second inequality follows directly from Ref.~\cite{cao2015discrete}.

Thus,
\begin{equation}
\begin{gathered}
\sqrt{Y_{k^a, k^b}^{\alpha, \mu} \, Y_{k^a, k^b}^{\alpha, \nu}}  + 
\sqrt{(1 - Y_{k^a, k^b}^{\alpha, \mu}) \, (1 - Y_{k^a, k^b}^{\alpha, \nu})} \\
\ge F_{\mu \nu} \\
\sqrt{Y_{k^a, k^b}^{\alpha, \mu} e_{k^a, k^b}^{\alpha, \mu} \, 
Y_{k^a, k^b}^{\alpha, \nu} e_{k^a, k^b}^{\alpha, \nu}}  + 
\\
\sqrt{(1 - Y_{k^a, k^b}^{\alpha, \mu} e_{k^a, k^b}^{\alpha, \mu}) \, 
(1 - Y_{k^a, k^b}^{\alpha, \nu} e_{k^a, k^b}^{\alpha, \nu})} 
\\
\ge F_{\mu \nu}
\end{gathered}
\end{equation}

From Ref.~\cite{cao2015discrete}, we can know that if
\begin{equation}
\sqrt{xy} + \sqrt{(1 - x) (1 - y)} \ge F_{\mu \nu},
\end{equation}
then
\begin{equation}
\left | x - y \right | \le \sqrt{1 - F_{\mu \nu}^{2}}.
\end{equation}

Finally, we can obtain
\begin{equation}
\begin{aligned}
\left | Y_{k^a, k^b}^{\alpha, \mu} - Y_{k^a, k^b}^{\alpha, \nu} \right | 
& \le \sqrt{1-F_{\mu\nu}^2},\\
\left | Y_{k^a, k^b}^{\alpha, \mu} e_{k^a, k^b}^{\alpha, \mu} - 
Y_{k^a, k^b}^{\alpha, \nu} e_{k^a, k^b}^{\alpha, \nu} \right | 
& \le \sqrt{1-F_{\mu\nu}^2}.
\end{aligned}
\end{equation}
This finishes the proof.

\section{Simulation} \label{App.C}
In this section, we present the detailed simulation formulas for the discrete-phase-randomized mode-pairing protocol.

Within the DPR-MP-QKD framework, the $Z$-basis is used to generate keys. Asymptotically, Alice and Bob primarily select intensities from $\left \{ 0, \mu \right \}$ for their signal pulses, each with probability approaching $1/2$, while the intensity $\nu$ for the decoy pulses is chosen only with negligible probability. In the $i$-th round ($i\in \mathrm{lim}_{N \to \infty }  \left \{ 1, 2, 3, ..., N \right \}$), Alice (Bob) prepares a coherent state $\big| \sqrt{z_i^a \mu} e^{i \phi _{i}^{a}} \big\rangle$ ($\big| \sqrt{z_i^b \mu} e^{i \phi _{i}^{b}} \big\rangle$) with $z_i^a$ $( z_i^b)$ $\in \left \{ 0, 1 \right \}$ and a phase $\phi _{i}^{a}$ $(\phi _{i}^{b})$ uniformly chosen from $D$ different values $\left \{ \frac{2\pi}{D}n \mid n\in 0, 1, 2, ..., D-1 \right \}$ (i.e., each with probability $1/D$). Here, $D$ represents the number of discrete phases, which is required to be an even integer. The $i$-th round intensity setting is then denoted by a 2-bit vector $z:=\left [ z_i^a, z_i^b \right ]$.

For the simulation, we assume that Alice and Bob transmit optical pulses to Charlie through a symmetrically attenuating channel. The single-side transmittance, which includes the detector efficiency $\eta_d$, is denoted as $\eta_s$, and each detector exhibits a dark count rate of $p_d$. The corresponding simulation parameters are listed in Table~\ref{tab:para}. In the settings of DPR-MP-QKD, the channel is independent and identically distributed (i.i.d.) for each round. Alice and Bob proceed to pair detection-clicked pulses and assign their measurement bases. For a candidate pulse pair $(i,j)$, define the parameter $\tau_{i,j} := \left[ \tau_{i,j}^a, \tau_{i,j}^b \right] = \left[ z_i^a \oplus z_j^a, z_i^b \oplus z_j^b \right]$, where $\oplus$ denotes bit-wise addition modulo 2. A signal pair is identified when $\tau_{i,j} = [1, 1]$.

Based on the key rate formula given by Eq.~(\ref{key rate fomula}), we first provide the expression for $r_p(p,l)$ as follows,
\begin{equation}
r_p(p,l) = \left [ \frac{1}{p\left [1 - \left ( 1 - p \right )^l \right ]} + 
\frac{1}{p} \right ] ^{-1},
\end{equation}
where $l$ is the maximal pairing interval and $p$ is the average click probability over all rounds, which will be discussed below. A detailed derivation of $r_p$ can refer to Ref.~\cite{zeng2022mode}.

The calculation of the parameters $r_s$, $E_{\mu,\mu}^Z$, and $q_{1,1}^Z$ later is based on the triggering effect of individual pulses. We denote the detection outcome in the $i$-th round as  $(L_i, R_i)$ for the left and right detectors, respectively. A successful click event is then defined as $C_i = L_i \oplus R_i = 1$. Thus, the detection probability $\mathrm{Pr}(C_i =1|z_i)$ is given by
\begin{equation}
\mathrm{Pr}(C_i = 1|z_i) \approx 1 - (1 - 2p_d) \, 
\mathrm{exp}\left [ -\eta_s \mu\left ( z_i^a + z_i^b \right )  \right ].
\end{equation}

The expected average click probability, i.e., the total transmittance of each round, is
\begin{equation}
\begin{aligned}
p & := \mathrm{Pr}(C_i = 1) = \sum_{z_i} \mathrm{Pr}(C_i = 1|z_i) \mathrm{Pr}(z_i) \\
& = \frac{1}{4} \mathrm{Pr}(C_i = 1|z_i).
\end{aligned}
\end{equation}

The discrete-phase-randomized coherent state emitted in the $i$-th round can be decomposed into pseudo photon number states as shown in Eq.~(\ref{pks}), and possesses an expected photon number of
\begin{equation}
\begin{aligned}
n_{k} & = \frac{1}{P _{k}^{\mu}} e^{-\mu} \sum_{m = 0}^{\infty}
\frac{\mu^{\left ( mD+k \right )}} {\left ( mD+k \right ) !} (mD+k) \\
& = \frac{\sum_{m = 0}^{\infty} \frac{\mu^{\left ( mD+k \right )}} {\left ( mD+k-1 \right ) !}}
{\sum_{l = 0}^{\infty} \frac{\mu^{\left ( lD+k \right )}} {\left ( lD+k \right ) !}} \\
& \approx k + \frac{\mu^D}{\frac{(D+k)!}{Dk!} + \frac{\mu^D}{D}},
\end{aligned}
\end{equation}
where the last equation follows from a first-order approximation. $\mathrm{Pr}(C_i=1|k_i)$ is defined as the joint detection probability with Alice and Bob, respectively, sending the pseudo photon number states $\big | \lambda_{k_i^a}^{\mu} \big \rangle$ and $\big | \lambda_{k_i^b}^{\mu} \big \rangle$, which is given by
\begin{equation}
\mathrm{Pr}(C_i = 1|k_i) \approx 1 - (1 - 2p_d)(1 - \eta_s)^{(n_{k_i^a} + n_{k_i^b})},
\label{cp_k}
\end{equation}
where we define $k_i  := [k_i^a, k_i^b]$.

We now proceed to calculate the signal-pair ratio $r_s$. Generally, we focus on the pair formed by the $i$-th and $j$-th rounds. For a general round $i$, the probability of intensity setting $z$ given a detection event is expressed as
\begin{equation}
\mathrm{Pr}(z_i|C = 1) = \frac{\mathrm{Pr}(z_i,C = 1)}{\mathrm{Pr}(C = 1)} 
= \frac{\mathrm{Pr}(C = 1|z_i)}{\sum_{z_i'} \mathrm{Pr}(C = 1|z_i')}.
\end{equation}

In our scheme, a successful click corresponds to $\tau _{i,j} = \left [ 1, 1 \right ]$, which arises from four possible configurations of $\left [ z_i, z_j \right ]$,
\begin{equation}
\left [ z_i, z_j \right ] \in \left \{ 
\left [ 00,11 \right ],
\left [ 01,10 \right ],
\left [ 10,01 \right ],
\left [ 11,00 \right ]
\right \},
\end{equation}
where $Err := \left \{\left [ 00,11 \right ], \left [ 11,00 \right ] \right \}$ denotes the set of bit error cases. For notational simplicity, we define the following events,
\begin{equation}
\scalebox{0.85}{$\displaystyle
\begin{aligned}
& \mathrm{Pr}(C) = \mathrm{Pr}(Pair \,  Clicked) := \mathrm{Pr}(C_i = C_j = 1) = p^2,  \\
& \mathrm{Pr}(E) = \mathrm{Pr}(Pair \,  Effective) :=\mathrm{Pr}(z_i \oplus z_j = 11), \\
& \mathrm{Pr}(Err) = \mathrm{Pr}(Pair \,  Erroneous) := \mathrm{Pr}(\left [ z_i, z_j \right] \in Err), \\
& \mathrm{Pr}(S) = \mathrm{Pr}(pseudo \, Single-photon \,  Pair  ) := \mathrm{Pr}(k_i \oplus k_j = 11).
\end{aligned}
$}
\end{equation}

The signal-pair ratio $r_s$ is given by
\begin{equation}
\begin{aligned}
r_s & = \mathrm{Pr}(E|C) = \mathrm{Pr}(z_i \oplus z_j = 11|C_i = 1, C_j = 1) \\
& = \sum_{z_i \oplus z_j = 11} \mathrm{Pr}(z_i = 1|C_i = 1)\mathrm{Pr}(z_j = 1|C_j = 1) \\
& = \sum_{z_i \oplus z_j = 11} \frac{\mathrm{Pr}(C_i = 1|z_i) \mathrm{Pr}(z_i)}
{\mathrm{Pr}(C_i = 1)} \frac{\mathrm{Pr}(C_j = 1|z_j) \mathrm{Pr}(z_j)}
{\mathrm{Pr}(C_j = 1)} \\
& = \frac{1}{16} \frac{1}{p^2} \sum_{z_i \oplus z_j = 11} 
\mathrm{Pr}(C_i = 1|z_i) \mathrm{Pr}(C_j = 1|z_j),
\end{aligned}
\end{equation}
when $\eta_s \mu \ll 1$, the signal-pair ratio $r_s$ is approximately
\begin{equation}
\begin{aligned}
r_s & \approx \frac{1}{16} \frac{1}{p^2} \left [ 2 \left ( \eta _s \mu  \right )^2 \right ] \\
& = \frac{1}{16} \frac{1}{\left ( \eta _s \mu  \right )^2} 
\left [ 2 \left ( \eta _s \mu  \right )^2 \right ] = \frac{1}{8},
\end{aligned}
\end{equation}
which is nearly a constant independent of $\eta_s$ and $\mu$.

The expected QBER $E_{\mu, \mu}^{Z}$ of the $(i, j)$-pair is
\begin{equation}
\begin{aligned}
& E_{\mu, \mu}^{Z} = \mathrm{Pr}(Err|E,C) \\
& = \frac{\mathrm{Pr}(Err,E|C)}{\mathrm{Pr}(E|C)} = 
\frac{\mathrm{Pr}(Err|C)}{\mathrm{Pr}(E|C)} \\
& = \frac{1}{r_s} \mathrm{Pr}(\left [ z_i, z_j \right ] \in Err | C_i = C_j = 1) \\
& = \frac{1}{r_s} \sum_{\left [ z_i, z_j \right ] \in Err} 
\mathrm{Pr}(z_i|C_i = 1) \mathrm{Pr}(z_j|C_j = 1) \\
& = \frac{1}{r_s} \sum_{\left [ z_i, z_j \right ] \in Err}
\frac{\mathrm{Pr}(C_i = 1|z_i)\mathrm{Pr}(z_i)}{\mathrm{Pr}(C_i = 1)}
\frac{\mathrm{Pr}(C_j = 1|z_j)\mathrm{Pr}(z_j)}{\mathrm{Pr}(C_j = 1)} \\
& = \frac{1}{16} \frac{1}{r_s p^2} \sum_{\left [ z_i, z_j \right ] \in Err}
\mathrm{Pr}(C_i = 1|z_i) \mathrm{Pr}(C_j = 1|z_j),
\end{aligned}
\end{equation}
where the third equality follows from the fact that the erroneous pair condition is a subset of the effective pair condition.

We then evaluate the expected pseudo single-photon pair ratio $q_{1,1}^Z$ within the effective signal pairs,
\begin{equation}
\begin{aligned}
q_{1,1}^Z & = \mathrm{Pr}(S|E,C) = \frac{\mathrm{Pr}(S,E,C)}{\mathrm{Pr}(E,C)} \\
& = \frac{1}{r_s p^2} \mathrm{Pr}(S,E,C) \\
& = \frac{1}{r_s p^2} \sum_{z_i,z_j} \mathrm{Pr}(S,E,C|z_i,z_j) \mathrm{Pr}(z_i,z_j) \\
& = \frac{1}{16} \frac{1}{r_s p^2} \sum_{z_i \oplus z_j = 11}
\mathrm{Pr}(S,C|z_i,z_j) \mathrm{Pr}(z_i,z_j) \\
& = \frac{1}{16} \frac{1}{r_s p^2} \sum_{z_i \oplus z_j = 11}
\mathrm{Pr}(C|S,z_i,z_j) \mathrm{Pr}(S|z_i,z_j) \\
& = \frac{1}{16} \frac{(P_1^{\mu})^2}{r_s p^2} \sum_{z_i \oplus z_j = 11}
\mathrm{Pr}(C_i = 1|k_i = z_i) \\
& \times \mathrm{Pr}(C_j = 1|k_j = z_j),
\end{aligned}
\end{equation}
where $P_k^{\mu}$ is given in Eq.~(\ref{dpr-pro}) and $\mathrm{Pr}(C_i = 1|k_i)$ is defined in Eq.~(\ref{cp_k}).

Since our simulation employs a single-pulse-based approach, we cannot directly calculate the phase error rate in the $Z$-basis under discrete phase randomization, consistent with the original MP-QKD protocol \cite{zeng2022mode}. Fortunately, if the decoy-state method is successfully implemented, we can first bound the bit error rate in the $X$-basis using the following formula \cite{ma2012alternative},
\begin{equation}
\begin{aligned}
Y_{1,1} & = (1-p_d^2) [ \frac{\eta _a \eta_b}{2} + 
(2\eta _a + 2\eta _b - 3\eta _a \eta _b)p_d \\
& + 4(1 - \eta _a)(1 - \eta _b)p_d^2 ], \\
e_{1,1} Y_{1,1} & = e_0 Y_{1,1} - (e_0 - e_d)(1 - p_d^2) \frac{\eta _a \eta_b}{2}, \\
& \left | Y_{1,1}^{2\mu,2\mu} - Y_{1,1}\right | \le \sqrt{1 - F_{\mu1}^2}, \\ 
& \left | Y_{1,1}^{2\mu,2\mu} e_{1,1}^{2\mu,2\mu} - Y_{1,1}e_{1,1}\right | \le \sqrt{1 - F_{\mu1}^2},
\end{aligned}
\end{equation}
where $e_0 = 1/2$ is the error rate for the vacuum, $e_d$ is the misalignment error, $Y_{1,1}^{2\mu,2\mu}$ and $e_{1,1}^{2\mu,2\mu}$ denote the yield and the bit error rate, respectively, for the $X$-basis in the DPR-MP-QKD protocol. The $F_{\mu1}$ is given by
\begin{equation}
F_{\mu1} := F(\big | \lambda_1^{2\mu} \big \rangle, \big | 1  \big \rangle ) = 
\frac{1}{\sqrt{\sum_{m=0}^{\infty}\frac{(2\mu)^{mD}}{(mD+1)!}}}.
\end{equation}
Once we obtain the upper and lower bounds for $Y_{1,1}^{2\mu,2\mu}$ and $e_{1,1}^{2\mu,2\mu}$, we can then utilize Eq.~(\ref{phase bound}) and Eq.~(\ref{dev}) to estimate the phase error rate $e_{1,1}^{Z,p}$ in the $Z$-basis.

\nocite{*}

\bibliography{DPR-MP-QKD}

\begin{thebibliography}{38}%
\makeatletter
\providecommand \@ifxundefined [1]{%
 \@ifx{#1\undefined}
}%
\providecommand \@ifnum [1]{%
 \ifnum #1\expandafter \@firstoftwo
 \else \expandafter \@secondoftwo
 \fi
}%
\providecommand \@ifx [1]{%
 \ifx #1\expandafter \@firstoftwo
 \else \expandafter \@secondoftwo
 \fi
}%
\providecommand \natexlab [1]{#1}%
\providecommand \enquote  [1]{``#1''}%
\providecommand \bibnamefont  [1]{#1}%
\providecommand \bibfnamefont [1]{#1}%
\providecommand \citenamefont [1]{#1}%
\providecommand \href@noop [0]{\@secondoftwo}%
\providecommand \href [0]{\begingroup \@sanitize@url \@href}%
\providecommand \@href[1]{\@@startlink{#1}\@@href}%
\providecommand \@@href[1]{\endgroup#1\@@endlink}%
\providecommand \@sanitize@url [0]{\catcode `\\12\catcode `\$12\catcode
  `\&12\catcode `\#12\catcode `\^12\catcode `\_12\catcode `\%12\relax}%
\providecommand \@@startlink[1]{}%
\providecommand \@@endlink[0]{}%
\providecommand \url  [0]{\begingroup\@sanitize@url \@url }%
\providecommand \@url [1]{\endgroup\@href {#1}{\urlprefix }}%
\providecommand \urlprefix  [0]{URL }%
\providecommand \Eprint [0]{\href }%
\providecommand \doibase [0]{https://doi.org/}%
\providecommand \selectlanguage [0]{\@gobble}%
\providecommand \bibinfo  [0]{\@secondoftwo}%
\providecommand \bibfield  [0]{\@secondoftwo}%
\providecommand \translation [1]{[#1]}%
\providecommand \BibitemOpen [0]{}%
\providecommand \bibitemStop [0]{}%
\providecommand \bibitemNoStop [0]{.\EOS\space}%
\providecommand \EOS [0]{\spacefactor3000\relax}%
\providecommand \BibitemShut  [1]{\csname bibitem#1\endcsname}%
\let\auto@bib@innerbib\@empty
\bibitem [{\citenamefont {Bennett}\ and\ \citenamefont
  {Brassard}(2014)}]{bennett2014quantum}%
  \BibitemOpen
  \bibfield  {author} {\bibinfo {author} {\bibfnamefont {C.~H.}\ \bibnamefont
  {Bennett}}\ and\ \bibinfo {author} {\bibfnamefont {G.}~\bibnamefont
  {Brassard}},\ }\bibfield  {title} {\bibinfo {title} {Quantum cryptography:
  Public key distribution and coin tossing},\ }\href@noop {} {\bibfield
  {journal} {\bibinfo  {journal} {Theoretical computer science}\ }\textbf
  {\bibinfo {volume} {560}},\ \bibinfo {pages} {7} (\bibinfo {year}
  {2014})}\BibitemShut {NoStop}%
\bibitem [{\citenamefont {Chen}\ \emph {et~al.}(2021)\citenamefont {Chen},
  \citenamefont {Zhang}, \citenamefont {Chen}, \citenamefont {Cai},
  \citenamefont {Liao}, \citenamefont {Zhang}, \citenamefont {Chen},
  \citenamefont {Yin}, \citenamefont {Ren}, \citenamefont {Chen} \emph
  {et~al.}}]{chen2021integrated}%
  \BibitemOpen
  \bibfield  {author} {\bibinfo {author} {\bibfnamefont {Y.-A.}\ \bibnamefont
  {Chen}}, \bibinfo {author} {\bibfnamefont {Q.}~\bibnamefont {Zhang}},
  \bibinfo {author} {\bibfnamefont {T.-Y.}\ \bibnamefont {Chen}}, \bibinfo
  {author} {\bibfnamefont {W.-Q.}\ \bibnamefont {Cai}}, \bibinfo {author}
  {\bibfnamefont {S.-K.}\ \bibnamefont {Liao}}, \bibinfo {author}
  {\bibfnamefont {J.}~\bibnamefont {Zhang}}, \bibinfo {author} {\bibfnamefont
  {K.}~\bibnamefont {Chen}}, \bibinfo {author} {\bibfnamefont {J.}~\bibnamefont
  {Yin}}, \bibinfo {author} {\bibfnamefont {J.-G.}\ \bibnamefont {Ren}},
  \bibinfo {author} {\bibfnamefont {Z.}~\bibnamefont {Chen}}, \emph {et~al.},\
  }\bibfield  {title} {\bibinfo {title} {An integrated space-to-ground quantum
  communication network over 4,600 kilometres},\ }\href@noop {} {\bibfield
  {journal} {\bibinfo  {journal} {Nature}\ }\textbf {\bibinfo {volume} {589}},\
  \bibinfo {pages} {214} (\bibinfo {year} {2021})}\BibitemShut {NoStop}%
\bibitem [{\citenamefont {Pirandola}\ \emph {et~al.}(2020)\citenamefont
  {Pirandola}, \citenamefont {Andersen}, \citenamefont {Banchi}, \citenamefont
  {Berta}, \citenamefont {Bunandar}, \citenamefont {Colbeck}, \citenamefont
  {Englund}, \citenamefont {Gehring}, \citenamefont {Lupo}, \citenamefont
  {Ottaviani} \emph {et~al.}}]{pirandola2020advances}%
  \BibitemOpen
  \bibfield  {author} {\bibinfo {author} {\bibfnamefont {S.}~\bibnamefont
  {Pirandola}}, \bibinfo {author} {\bibfnamefont {U.~L.}\ \bibnamefont
  {Andersen}}, \bibinfo {author} {\bibfnamefont {L.}~\bibnamefont {Banchi}},
  \bibinfo {author} {\bibfnamefont {M.}~\bibnamefont {Berta}}, \bibinfo
  {author} {\bibfnamefont {D.}~\bibnamefont {Bunandar}}, \bibinfo {author}
  {\bibfnamefont {R.}~\bibnamefont {Colbeck}}, \bibinfo {author} {\bibfnamefont
  {D.}~\bibnamefont {Englund}}, \bibinfo {author} {\bibfnamefont
  {T.}~\bibnamefont {Gehring}}, \bibinfo {author} {\bibfnamefont
  {C.}~\bibnamefont {Lupo}}, \bibinfo {author} {\bibfnamefont {C.}~\bibnamefont
  {Ottaviani}}, \emph {et~al.},\ }\bibfield  {title} {\bibinfo {title}
  {Advances in quantum cryptography},\ }\href@noop {} {\bibfield  {journal}
  {\bibinfo  {journal} {Advances in optics and photonics}\ }\textbf {\bibinfo
  {volume} {12}},\ \bibinfo {pages} {1012} (\bibinfo {year}
  {2020})}\BibitemShut {NoStop}%
\bibitem [{\citenamefont {Lydersen}\ \emph {et~al.}(2010)\citenamefont
  {Lydersen}, \citenamefont {Wiechers}, \citenamefont {Wittmann}, \citenamefont
  {Elser}, \citenamefont {Skaar},\ and\ \citenamefont
  {Makarov}}]{lydersen2010hacking}%
  \BibitemOpen
  \bibfield  {author} {\bibinfo {author} {\bibfnamefont {L.}~\bibnamefont
  {Lydersen}}, \bibinfo {author} {\bibfnamefont {C.}~\bibnamefont {Wiechers}},
  \bibinfo {author} {\bibfnamefont {C.}~\bibnamefont {Wittmann}}, \bibinfo
  {author} {\bibfnamefont {D.}~\bibnamefont {Elser}}, \bibinfo {author}
  {\bibfnamefont {J.}~\bibnamefont {Skaar}},\ and\ \bibinfo {author}
  {\bibfnamefont {V.}~\bibnamefont {Makarov}},\ }\bibfield  {title} {\bibinfo
  {title} {Hacking commercial quantum cryptography systems by tailored bright
  illumination},\ }\href@noop {} {\bibfield  {journal} {\bibinfo  {journal}
  {Nature photonics}\ }\textbf {\bibinfo {volume} {4}},\ \bibinfo {pages} {686}
  (\bibinfo {year} {2010})}\BibitemShut {NoStop}%
\bibitem [{\citenamefont {Xu}\ \emph {et~al.}(2020)\citenamefont {Xu},
  \citenamefont {Ma}, \citenamefont {Zhang}, \citenamefont {Lo},\ and\
  \citenamefont {Pan}}]{xu2020secure}%
  \BibitemOpen
  \bibfield  {author} {\bibinfo {author} {\bibfnamefont {F.}~\bibnamefont
  {Xu}}, \bibinfo {author} {\bibfnamefont {X.}~\bibnamefont {Ma}}, \bibinfo
  {author} {\bibfnamefont {Q.}~\bibnamefont {Zhang}}, \bibinfo {author}
  {\bibfnamefont {H.-K.}\ \bibnamefont {Lo}},\ and\ \bibinfo {author}
  {\bibfnamefont {J.-W.}\ \bibnamefont {Pan}},\ }\bibfield  {title} {\bibinfo
  {title} {Secure quantum key distribution with realistic devices},\
  }\href@noop {} {\bibfield  {journal} {\bibinfo  {journal} {Reviews of modern
  physics}\ }\textbf {\bibinfo {volume} {92}},\ \bibinfo {pages} {025002}
  (\bibinfo {year} {2020})}\BibitemShut {NoStop}%
\bibitem [{\citenamefont {Takeoka}\ \emph {et~al.}(2014)\citenamefont
  {Takeoka}, \citenamefont {Guha},\ and\ \citenamefont
  {Wilde}}]{takeoka2014fundamental}%
  \BibitemOpen
  \bibfield  {author} {\bibinfo {author} {\bibfnamefont {M.}~\bibnamefont
  {Takeoka}}, \bibinfo {author} {\bibfnamefont {S.}~\bibnamefont {Guha}},\ and\
  \bibinfo {author} {\bibfnamefont {M.~M.}\ \bibnamefont {Wilde}},\ }\bibfield
  {title} {\bibinfo {title} {Fundamental rate-loss tradeoff for optical quantum
  key distribution},\ }\href@noop {} {\bibfield  {journal} {\bibinfo  {journal}
  {Nature communications}\ }\textbf {\bibinfo {volume} {5}},\ \bibinfo {pages}
  {5235} (\bibinfo {year} {2014})}\BibitemShut {NoStop}%
\bibitem [{\citenamefont {Pirandola}\ \emph {et~al.}(2017)\citenamefont
  {Pirandola}, \citenamefont {Laurenza}, \citenamefont {Ottaviani},\ and\
  \citenamefont {Banchi}}]{pirandola2017fundamental}%
  \BibitemOpen
  \bibfield  {author} {\bibinfo {author} {\bibfnamefont {S.}~\bibnamefont
  {Pirandola}}, \bibinfo {author} {\bibfnamefont {R.}~\bibnamefont {Laurenza}},
  \bibinfo {author} {\bibfnamefont {C.}~\bibnamefont {Ottaviani}},\ and\
  \bibinfo {author} {\bibfnamefont {L.}~\bibnamefont {Banchi}},\ }\bibfield
  {title} {\bibinfo {title} {Fundamental limits of repeaterless quantum
  communications},\ }\href@noop {} {\bibfield  {journal} {\bibinfo  {journal}
  {Nature communications}\ }\textbf {\bibinfo {volume} {8}},\ \bibinfo {pages}
  {15043} (\bibinfo {year} {2017})}\BibitemShut {NoStop}%
\bibitem [{\citenamefont {Lo}\ and\ \citenamefont
  {Chau}(1999)}]{lo1999unconditional}%
  \BibitemOpen
  \bibfield  {author} {\bibinfo {author} {\bibfnamefont {H.-K.}\ \bibnamefont
  {Lo}}\ and\ \bibinfo {author} {\bibfnamefont {H.~F.}\ \bibnamefont {Chau}},\
  }\bibfield  {title} {\bibinfo {title} {Unconditional security of quantum key
  distribution over arbitrarily long distances},\ }\href@noop {} {\bibfield
  {journal} {\bibinfo  {journal} {science}\ }\textbf {\bibinfo {volume}
  {283}},\ \bibinfo {pages} {2050} (\bibinfo {year} {1999})}\BibitemShut
  {NoStop}%
\bibitem [{\citenamefont {Shor}\ and\ \citenamefont
  {Preskill}(2000)}]{shor2000simple}%
  \BibitemOpen
  \bibfield  {author} {\bibinfo {author} {\bibfnamefont {P.~W.}\ \bibnamefont
  {Shor}}\ and\ \bibinfo {author} {\bibfnamefont {J.}~\bibnamefont
  {Preskill}},\ }\bibfield  {title} {\bibinfo {title} {Simple proof of security
  of the bb84 quantum key distribution protocol},\ }\href@noop {} {\bibfield
  {journal} {\bibinfo  {journal} {Physical review letters}\ }\textbf {\bibinfo
  {volume} {85}},\ \bibinfo {pages} {441} (\bibinfo {year} {2000})}\BibitemShut
  {NoStop}%
\bibitem [{\citenamefont {Koashi}(2009)}]{koashi2009simple}%
  \BibitemOpen
  \bibfield  {author} {\bibinfo {author} {\bibfnamefont {M.}~\bibnamefont
  {Koashi}},\ }\bibfield  {title} {\bibinfo {title} {Simple security proof of
  quantum key distribution based on complementarity},\ }\href@noop {}
  {\bibfield  {journal} {\bibinfo  {journal} {New Journal of Physics}\ }\textbf
  {\bibinfo {volume} {11}},\ \bibinfo {pages} {045018} (\bibinfo {year}
  {2009})}\BibitemShut {NoStop}%
\bibitem [{\citenamefont {Lo}\ and\ \citenamefont
  {Preskill}(2007)}]{lo2007security}%
  \BibitemOpen
  \bibfield  {author} {\bibinfo {author} {\bibfnamefont {H.~K.}\ \bibnamefont
  {Lo}}\ and\ \bibinfo {author} {\bibfnamefont {J.}~\bibnamefont {Preskill}},\
  }\bibfield  {title} {\bibinfo {title} {Security of quantum key distribution
  using weak coherent states with nonrandom phases},\ }\href@noop {} {\bibfield
   {journal} {\bibinfo  {journal} {Quantum Information and Computation}\ }
  (\bibinfo {year} {2007})}\BibitemShut {NoStop}%
\bibitem [{\citenamefont {Lo}\ \emph {et~al.}(2012)\citenamefont {Lo},
  \citenamefont {Curty},\ and\ \citenamefont {Qi}}]{lo2012measurement}%
  \BibitemOpen
  \bibfield  {author} {\bibinfo {author} {\bibfnamefont {H.-K.}\ \bibnamefont
  {Lo}}, \bibinfo {author} {\bibfnamefont {M.}~\bibnamefont {Curty}},\ and\
  \bibinfo {author} {\bibfnamefont {B.}~\bibnamefont {Qi}},\ }\bibfield
  {title} {\bibinfo {title} {Measurement-device-independent quantum key
  distribution},\ }\href@noop {} {\bibfield  {journal} {\bibinfo  {journal}
  {Physical review letters}\ }\textbf {\bibinfo {volume} {108}},\ \bibinfo
  {pages} {130503} (\bibinfo {year} {2012})}\BibitemShut {NoStop}%
\bibitem [{\citenamefont {Tamaki}\ \emph {et~al.}(2012)\citenamefont {Tamaki},
  \citenamefont {Lo}, \citenamefont {Fung},\ and\ \citenamefont
  {Qi}}]{tamaki2012phase}%
  \BibitemOpen
  \bibfield  {author} {\bibinfo {author} {\bibfnamefont {K.}~\bibnamefont
  {Tamaki}}, \bibinfo {author} {\bibfnamefont {H.-K.}\ \bibnamefont {Lo}},
  \bibinfo {author} {\bibfnamefont {C.-H.~F.}\ \bibnamefont {Fung}},\ and\
  \bibinfo {author} {\bibfnamefont {B.}~\bibnamefont {Qi}},\ }\bibfield
  {title} {\bibinfo {title} {Phase encoding schemes for
  measurement-device-independent quantum key distribution with basis-dependent
  flaw},\ }\href@noop {} {\bibfield  {journal} {\bibinfo  {journal} {Physical
  Review A—Atomic, Molecular, and Optical Physics}\ }\textbf {\bibinfo
  {volume} {85}},\ \bibinfo {pages} {042307} (\bibinfo {year}
  {2012})}\BibitemShut {NoStop}%
\bibitem [{\citenamefont {Ma}\ and\ \citenamefont
  {Razavi}(2012)}]{ma2012alternative}%
  \BibitemOpen
  \bibfield  {author} {\bibinfo {author} {\bibfnamefont {X.}~\bibnamefont
  {Ma}}\ and\ \bibinfo {author} {\bibfnamefont {M.}~\bibnamefont {Razavi}},\
  }\bibfield  {title} {\bibinfo {title} {Alternative schemes for
  measurement-device-independent quantum key distribution},\ }\href@noop {}
  {\bibfield  {journal} {\bibinfo  {journal} {Physical Review A—Atomic,
  Molecular, and Optical Physics}\ }\textbf {\bibinfo {volume} {86}},\ \bibinfo
  {pages} {062319} (\bibinfo {year} {2012})}\BibitemShut {NoStop}%
\bibitem [{\citenamefont {Liu}\ \emph {et~al.}(2013)\citenamefont {Liu},
  \citenamefont {Chen}, \citenamefont {Wang}, \citenamefont {Liang},
  \citenamefont {Shentu}, \citenamefont {Wang}, \citenamefont {Cui},
  \citenamefont {Yin}, \citenamefont {Liu}, \citenamefont {Li} \emph
  {et~al.}}]{liu2013experimental}%
  \BibitemOpen
  \bibfield  {author} {\bibinfo {author} {\bibfnamefont {Y.}~\bibnamefont
  {Liu}}, \bibinfo {author} {\bibfnamefont {T.-Y.}\ \bibnamefont {Chen}},
  \bibinfo {author} {\bibfnamefont {L.-J.}\ \bibnamefont {Wang}}, \bibinfo
  {author} {\bibfnamefont {H.}~\bibnamefont {Liang}}, \bibinfo {author}
  {\bibfnamefont {G.-L.}\ \bibnamefont {Shentu}}, \bibinfo {author}
  {\bibfnamefont {J.}~\bibnamefont {Wang}}, \bibinfo {author} {\bibfnamefont
  {K.}~\bibnamefont {Cui}}, \bibinfo {author} {\bibfnamefont {H.-L.}\
  \bibnamefont {Yin}}, \bibinfo {author} {\bibfnamefont {N.-L.}\ \bibnamefont
  {Liu}}, \bibinfo {author} {\bibfnamefont {L.}~\bibnamefont {Li}}, \emph
  {et~al.},\ }\bibfield  {title} {\bibinfo {title} {Experimental
  measurement-device-independent quantum key distribution},\ }\href@noop {}
  {\bibfield  {journal} {\bibinfo  {journal} {Physical review letters}\
  }\textbf {\bibinfo {volume} {111}},\ \bibinfo {pages} {130502} (\bibinfo
  {year} {2013})}\BibitemShut {NoStop}%
\bibitem [{\citenamefont {Ferreira~da Silva}\ \emph {et~al.}(2013)\citenamefont
  {Ferreira~da Silva}, \citenamefont {Vitoreti}, \citenamefont {Xavier},
  \citenamefont {Do~Amaral}, \citenamefont {Tempor{\~a}o},\ and\ \citenamefont
  {Von Der~Weid}}]{ferreira2013proof}%
  \BibitemOpen
  \bibfield  {author} {\bibinfo {author} {\bibfnamefont {T.}~\bibnamefont
  {Ferreira~da Silva}}, \bibinfo {author} {\bibfnamefont {D.}~\bibnamefont
  {Vitoreti}}, \bibinfo {author} {\bibfnamefont {G.}~\bibnamefont {Xavier}},
  \bibinfo {author} {\bibfnamefont {G.}~\bibnamefont {Do~Amaral}}, \bibinfo
  {author} {\bibfnamefont {G.}~\bibnamefont {Tempor{\~a}o}},\ and\ \bibinfo
  {author} {\bibfnamefont {J.}~\bibnamefont {Von Der~Weid}},\ }\bibfield
  {title} {\bibinfo {title} {Proof-of-principle demonstration of
  measurement-device-independent quantum key distribution using polarization
  qubits},\ }\href@noop {} {\bibfield  {journal} {\bibinfo  {journal} {Physical
  Review A—Atomic, Molecular, and Optical Physics}\ }\textbf {\bibinfo
  {volume} {88}},\ \bibinfo {pages} {052303} (\bibinfo {year}
  {2013})}\BibitemShut {NoStop}%
\bibitem [{\citenamefont {Tang}\ \emph {et~al.}(2014)\citenamefont {Tang},
  \citenamefont {Liao}, \citenamefont {Xu}, \citenamefont {Qi}, \citenamefont
  {Qian},\ and\ \citenamefont {Lo}}]{tang2014experimental}%
  \BibitemOpen
  \bibfield  {author} {\bibinfo {author} {\bibfnamefont {Z.}~\bibnamefont
  {Tang}}, \bibinfo {author} {\bibfnamefont {Z.}~\bibnamefont {Liao}}, \bibinfo
  {author} {\bibfnamefont {F.}~\bibnamefont {Xu}}, \bibinfo {author}
  {\bibfnamefont {B.}~\bibnamefont {Qi}}, \bibinfo {author} {\bibfnamefont
  {L.}~\bibnamefont {Qian}},\ and\ \bibinfo {author} {\bibfnamefont {H.-K.}\
  \bibnamefont {Lo}},\ }\bibfield  {title} {\bibinfo {title} {Experimental
  demonstration of polarization encoding measurement-device-independent quantum
  key distribution},\ }\href@noop {} {\bibfield  {journal} {\bibinfo  {journal}
  {Physical review letters}\ }\textbf {\bibinfo {volume} {112}},\ \bibinfo
  {pages} {190503} (\bibinfo {year} {2014})}\BibitemShut {NoStop}%
\bibitem [{\citenamefont {Woodward}\ \emph {et~al.}(2021)\citenamefont
  {Woodward}, \citenamefont {Lo}, \citenamefont {Pittaluga}, \citenamefont
  {Minder}, \citenamefont {Para{\"\i}so}, \citenamefont {Lucamarini},
  \citenamefont {Yuan},\ and\ \citenamefont {Shields}}]{woodward2021gigahertz}%
  \BibitemOpen
  \bibfield  {author} {\bibinfo {author} {\bibfnamefont {R.~I.}\ \bibnamefont
  {Woodward}}, \bibinfo {author} {\bibfnamefont {Y.}~\bibnamefont {Lo}},
  \bibinfo {author} {\bibfnamefont {M.}~\bibnamefont {Pittaluga}}, \bibinfo
  {author} {\bibfnamefont {M.}~\bibnamefont {Minder}}, \bibinfo {author}
  {\bibfnamefont {T.~K.}\ \bibnamefont {Para{\"\i}so}}, \bibinfo {author}
  {\bibfnamefont {M.}~\bibnamefont {Lucamarini}}, \bibinfo {author}
  {\bibfnamefont {Z.}~\bibnamefont {Yuan}},\ and\ \bibinfo {author}
  {\bibfnamefont {A.}~\bibnamefont {Shields}},\ }\bibfield  {title} {\bibinfo
  {title} {Gigahertz measurement-device-independent quantum key distribution
  using directly modulated lasers},\ }\href@noop {} {\bibfield  {journal}
  {\bibinfo  {journal} {npj Quantum Information}\ }\textbf {\bibinfo {volume}
  {7}},\ \bibinfo {pages} {58} (\bibinfo {year} {2021})}\BibitemShut {NoStop}%
\bibitem [{\citenamefont {Tang}\ \emph {et~al.}(2016)\citenamefont {Tang},
  \citenamefont {Yin}, \citenamefont {Zhao}, \citenamefont {Liu}, \citenamefont
  {Sun}, \citenamefont {Huang}, \citenamefont {Zhang}, \citenamefont {Chen},
  \citenamefont {Zhang}, \citenamefont {You} \emph
  {et~al.}}]{tang2016measurement}%
  \BibitemOpen
  \bibfield  {author} {\bibinfo {author} {\bibfnamefont {Y.-L.}\ \bibnamefont
  {Tang}}, \bibinfo {author} {\bibfnamefont {H.-L.}\ \bibnamefont {Yin}},
  \bibinfo {author} {\bibfnamefont {Q.}~\bibnamefont {Zhao}}, \bibinfo {author}
  {\bibfnamefont {H.}~\bibnamefont {Liu}}, \bibinfo {author} {\bibfnamefont
  {X.-X.}\ \bibnamefont {Sun}}, \bibinfo {author} {\bibfnamefont {M.-Q.}\
  \bibnamefont {Huang}}, \bibinfo {author} {\bibfnamefont {W.-J.}\ \bibnamefont
  {Zhang}}, \bibinfo {author} {\bibfnamefont {S.-J.}\ \bibnamefont {Chen}},
  \bibinfo {author} {\bibfnamefont {L.}~\bibnamefont {Zhang}}, \bibinfo
  {author} {\bibfnamefont {L.-X.}\ \bibnamefont {You}}, \emph {et~al.},\
  }\bibfield  {title} {\bibinfo {title} {Measurement-device-independent quantum
  key distribution over untrustful metropolitan network},\ }\href@noop {}
  {\bibfield  {journal} {\bibinfo  {journal} {Physical Review X}\ }\textbf
  {\bibinfo {volume} {6}},\ \bibinfo {pages} {011024} (\bibinfo {year}
  {2016})}\BibitemShut {NoStop}%
\bibitem [{\citenamefont {Zukowski}\ \emph {et~al.}(1993)\citenamefont
  {Zukowski}, \citenamefont {Zeilinger}, \citenamefont {Horne},\ and\
  \citenamefont {Ekert}}]{zukowski1993event}%
  \BibitemOpen
  \bibfield  {author} {\bibinfo {author} {\bibfnamefont {M.}~\bibnamefont
  {Zukowski}}, \bibinfo {author} {\bibfnamefont {A.}~\bibnamefont {Zeilinger}},
  \bibinfo {author} {\bibfnamefont {M.}~\bibnamefont {Horne}},\ and\ \bibinfo
  {author} {\bibfnamefont {A.}~\bibnamefont {Ekert}},\ }\bibfield  {title}
  {\bibinfo {title} {" event-ready-detectors" bell experiment via entanglement
  swapping.},\ }\href@noop {} {\bibfield  {journal} {\bibinfo  {journal}
  {Physical review letters}\ }\textbf {\bibinfo {volume} {71}} (\bibinfo {year}
  {1993})}\BibitemShut {NoStop}%
\bibitem [{\citenamefont {Briegel}\ \emph {et~al.}(1998)\citenamefont
  {Briegel}, \citenamefont {D{\"u}r}, \citenamefont {Cirac},\ and\
  \citenamefont {Zoller}}]{briegel1998quantum}%
  \BibitemOpen
  \bibfield  {author} {\bibinfo {author} {\bibfnamefont {H.-J.}\ \bibnamefont
  {Briegel}}, \bibinfo {author} {\bibfnamefont {W.}~\bibnamefont {D{\"u}r}},
  \bibinfo {author} {\bibfnamefont {J.~I.}\ \bibnamefont {Cirac}},\ and\
  \bibinfo {author} {\bibfnamefont {P.}~\bibnamefont {Zoller}},\ }\bibfield
  {title} {\bibinfo {title} {Quantum repeaters: the role of imperfect local
  operations in quantum communication},\ }\href@noop {} {\bibfield  {journal}
  {\bibinfo  {journal} {Physical Review Letters}\ }\textbf {\bibinfo {volume}
  {81}},\ \bibinfo {pages} {5932} (\bibinfo {year} {1998})}\BibitemShut
  {NoStop}%
\bibitem [{\citenamefont {Azuma}\ \emph {et~al.}(2015)\citenamefont {Azuma},
  \citenamefont {Tamaki},\ and\ \citenamefont {Lo}}]{azuma2015all}%
  \BibitemOpen
  \bibfield  {author} {\bibinfo {author} {\bibfnamefont {K.}~\bibnamefont
  {Azuma}}, \bibinfo {author} {\bibfnamefont {K.}~\bibnamefont {Tamaki}},\ and\
  \bibinfo {author} {\bibfnamefont {H.-K.}\ \bibnamefont {Lo}},\ }\bibfield
  {title} {\bibinfo {title} {All-photonic quantum repeaters},\ }\href@noop {}
  {\bibfield  {journal} {\bibinfo  {journal} {Nature communications}\ }\textbf
  {\bibinfo {volume} {6}},\ \bibinfo {pages} {6787} (\bibinfo {year}
  {2015})}\BibitemShut {NoStop}%
\bibitem [{\citenamefont {Lucamarini}\ \emph {et~al.}(2018)\citenamefont
  {Lucamarini}, \citenamefont {Yuan}, \citenamefont {Dynes},\ and\
  \citenamefont {Shields}}]{lucamarini2018overcoming}%
  \BibitemOpen
  \bibfield  {author} {\bibinfo {author} {\bibfnamefont {M.}~\bibnamefont
  {Lucamarini}}, \bibinfo {author} {\bibfnamefont {Z.~L.}\ \bibnamefont
  {Yuan}}, \bibinfo {author} {\bibfnamefont {J.~F.}\ \bibnamefont {Dynes}},\
  and\ \bibinfo {author} {\bibfnamefont {A.~J.}\ \bibnamefont {Shields}},\
  }\bibfield  {title} {\bibinfo {title} {Overcoming the rate--distance limit of
  quantum key distribution without quantum repeaters},\ }\href@noop {}
  {\bibfield  {journal} {\bibinfo  {journal} {Nature}\ }\textbf {\bibinfo
  {volume} {557}},\ \bibinfo {pages} {400} (\bibinfo {year}
  {2018})}\BibitemShut {NoStop}%
\bibitem [{\citenamefont {Ma}\ \emph {et~al.}(2018)\citenamefont {Ma},
  \citenamefont {Zeng},\ and\ \citenamefont {Zhou}}]{ma2018phase}%
  \BibitemOpen
  \bibfield  {author} {\bibinfo {author} {\bibfnamefont {X.}~\bibnamefont
  {Ma}}, \bibinfo {author} {\bibfnamefont {P.}~\bibnamefont {Zeng}},\ and\
  \bibinfo {author} {\bibfnamefont {H.}~\bibnamefont {Zhou}},\ }\bibfield
  {title} {\bibinfo {title} {Phase-matching quantum key distribution},\
  }\href@noop {} {\bibfield  {journal} {\bibinfo  {journal} {Physical Review
  X}\ }\textbf {\bibinfo {volume} {8}},\ \bibinfo {pages} {031043} (\bibinfo
  {year} {2018})}\BibitemShut {NoStop}%
\bibitem [{\citenamefont {Lin}\ and\ \citenamefont
  {L{\"u}tkenhaus}(2018)}]{lin2018simple}%
  \BibitemOpen
  \bibfield  {author} {\bibinfo {author} {\bibfnamefont {J.}~\bibnamefont
  {Lin}}\ and\ \bibinfo {author} {\bibfnamefont {N.}~\bibnamefont
  {L{\"u}tkenhaus}},\ }\bibfield  {title} {\bibinfo {title} {Simple security
  analysis of phase-matching measurement-device-independent quantum key
  distribution},\ }\href@noop {} {\bibfield  {journal} {\bibinfo  {journal}
  {Physical Review A}\ }\textbf {\bibinfo {volume} {98}},\ \bibinfo {pages}
  {042332} (\bibinfo {year} {2018})}\BibitemShut {NoStop}%
\bibitem [{\citenamefont {Wang}\ \emph {et~al.}(2018)\citenamefont {Wang},
  \citenamefont {Yu},\ and\ \citenamefont {Hu}}]{wang2018twin}%
  \BibitemOpen
  \bibfield  {author} {\bibinfo {author} {\bibfnamefont {X.-B.}\ \bibnamefont
  {Wang}}, \bibinfo {author} {\bibfnamefont {Z.-W.}\ \bibnamefont {Yu}},\ and\
  \bibinfo {author} {\bibfnamefont {X.-L.}\ \bibnamefont {Hu}},\ }\bibfield
  {title} {\bibinfo {title} {Twin-field quantum key distribution with large
  misalignment error},\ }\href@noop {} {\bibfield  {journal} {\bibinfo
  {journal} {Physical Review A}\ }\textbf {\bibinfo {volume} {98}},\ \bibinfo
  {pages} {062323} (\bibinfo {year} {2018})}\BibitemShut {NoStop}%
\bibitem [{\citenamefont {Zeng}\ \emph {et~al.}(2022)\citenamefont {Zeng},
  \citenamefont {Zhou}, \citenamefont {Wu},\ and\ \citenamefont
  {Ma}}]{zeng2022mode}%
  \BibitemOpen
  \bibfield  {author} {\bibinfo {author} {\bibfnamefont {P.}~\bibnamefont
  {Zeng}}, \bibinfo {author} {\bibfnamefont {H.}~\bibnamefont {Zhou}}, \bibinfo
  {author} {\bibfnamefont {W.}~\bibnamefont {Wu}},\ and\ \bibinfo {author}
  {\bibfnamefont {X.}~\bibnamefont {Ma}},\ }\bibfield  {title} {\bibinfo
  {title} {Mode-pairing quantum key distribution},\ }\href@noop {} {\bibfield
  {journal} {\bibinfo  {journal} {Nature Communications}\ }\textbf {\bibinfo
  {volume} {13}},\ \bibinfo {pages} {3903} (\bibinfo {year}
  {2022})}\BibitemShut {NoStop}%
\bibitem [{\citenamefont {Zhu}\ \emph {et~al.}(2023)\citenamefont {Zhu},
  \citenamefont {Huang}, \citenamefont {Liu}, \citenamefont {Zeng},
  \citenamefont {Zou}, \citenamefont {Dai}, \citenamefont {Tang}, \citenamefont
  {Li}, \citenamefont {You}, \citenamefont {Wang} \emph
  {et~al.}}]{zhu2023experimental}%
  \BibitemOpen
  \bibfield  {author} {\bibinfo {author} {\bibfnamefont {H.-T.}\ \bibnamefont
  {Zhu}}, \bibinfo {author} {\bibfnamefont {Y.}~\bibnamefont {Huang}}, \bibinfo
  {author} {\bibfnamefont {H.}~\bibnamefont {Liu}}, \bibinfo {author}
  {\bibfnamefont {P.}~\bibnamefont {Zeng}}, \bibinfo {author} {\bibfnamefont
  {M.}~\bibnamefont {Zou}}, \bibinfo {author} {\bibfnamefont {Y.}~\bibnamefont
  {Dai}}, \bibinfo {author} {\bibfnamefont {S.}~\bibnamefont {Tang}}, \bibinfo
  {author} {\bibfnamefont {H.}~\bibnamefont {Li}}, \bibinfo {author}
  {\bibfnamefont {L.}~\bibnamefont {You}}, \bibinfo {author} {\bibfnamefont
  {Z.}~\bibnamefont {Wang}}, \emph {et~al.},\ }\bibfield  {title} {\bibinfo
  {title} {Experimental mode-pairing measurement-device-independent quantum key
  distribution without global phase locking},\ }\href@noop {} {\bibfield
  {journal} {\bibinfo  {journal} {Physical Review Letters}\ }\textbf {\bibinfo
  {volume} {130}},\ \bibinfo {pages} {030801} (\bibinfo {year}
  {2023})}\BibitemShut {NoStop}%
\bibitem [{\citenamefont {Sun}\ \emph {et~al.}(2012)\citenamefont {Sun},
  \citenamefont {Gao}, \citenamefont {Jiang}, \citenamefont {Li},\ and\
  \citenamefont {Liang}}]{sun2012partially}%
  \BibitemOpen
  \bibfield  {author} {\bibinfo {author} {\bibfnamefont {S.-H.}\ \bibnamefont
  {Sun}}, \bibinfo {author} {\bibfnamefont {M.}~\bibnamefont {Gao}}, \bibinfo
  {author} {\bibfnamefont {M.-S.}\ \bibnamefont {Jiang}}, \bibinfo {author}
  {\bibfnamefont {C.-Y.}\ \bibnamefont {Li}},\ and\ \bibinfo {author}
  {\bibfnamefont {L.-M.}\ \bibnamefont {Liang}},\ }\bibfield  {title} {\bibinfo
  {title} {Partially random phase attack to the practical two-way
  quantum-key-distribution system},\ }\href
  {https://doi.org/10.1103/PhysRevA.85.032304} {\bibfield  {journal} {\bibinfo
  {journal} {Phys. Rev. A}\ }\textbf {\bibinfo {volume} {85}},\ \bibinfo
  {pages} {032304} (\bibinfo {year} {2012})}\BibitemShut {NoStop}%
\bibitem [{\citenamefont {Tang}\ \emph {et~al.}(2013)\citenamefont {Tang},
  \citenamefont {Yin}, \citenamefont {Ma}, \citenamefont {Fung}, \citenamefont
  {Liu}, \citenamefont {Yong}, \citenamefont {Chen}, \citenamefont {Peng},
  \citenamefont {Chen},\ and\ \citenamefont {Pan}}]{tang2013source}%
  \BibitemOpen
  \bibfield  {author} {\bibinfo {author} {\bibfnamefont {Y.-L.}\ \bibnamefont
  {Tang}}, \bibinfo {author} {\bibfnamefont {H.-L.}\ \bibnamefont {Yin}},
  \bibinfo {author} {\bibfnamefont {X.}~\bibnamefont {Ma}}, \bibinfo {author}
  {\bibfnamefont {C.-H.~F.}\ \bibnamefont {Fung}}, \bibinfo {author}
  {\bibfnamefont {Y.}~\bibnamefont {Liu}}, \bibinfo {author} {\bibfnamefont
  {H.-L.}\ \bibnamefont {Yong}}, \bibinfo {author} {\bibfnamefont {T.-Y.}\
  \bibnamefont {Chen}}, \bibinfo {author} {\bibfnamefont {C.-Z.}\ \bibnamefont
  {Peng}}, \bibinfo {author} {\bibfnamefont {Z.-B.}\ \bibnamefont {Chen}},\
  and\ \bibinfo {author} {\bibfnamefont {J.-W.}\ \bibnamefont {Pan}},\
  }\bibfield  {title} {\bibinfo {title} {Source attack of decoy-state quantum
  key distribution using phase information},\ }\href@noop {} {\bibfield
  {journal} {\bibinfo  {journal} {Physical Review A—Atomic, Molecular, and
  Optical Physics}\ }\textbf {\bibinfo {volume} {88}},\ \bibinfo {pages}
  {022308} (\bibinfo {year} {2013})}\BibitemShut {NoStop}%
\bibitem [{\citenamefont {Cao}\ \emph {et~al.}(2015)\citenamefont {Cao},
  \citenamefont {Zhang}, \citenamefont {Lo},\ and\ \citenamefont
  {Ma}}]{cao2015discrete}%
  \BibitemOpen
  \bibfield  {author} {\bibinfo {author} {\bibfnamefont {Z.}~\bibnamefont
  {Cao}}, \bibinfo {author} {\bibfnamefont {Z.}~\bibnamefont {Zhang}}, \bibinfo
  {author} {\bibfnamefont {H.-K.}\ \bibnamefont {Lo}},\ and\ \bibinfo {author}
  {\bibfnamefont {X.}~\bibnamefont {Ma}},\ }\bibfield  {title} {\bibinfo
  {title} {Discrete-phase-randomized coherent state source and its application
  in quantum key distribution},\ }\href@noop {} {\bibfield  {journal} {\bibinfo
   {journal} {New Journal of Physics}\ }\textbf {\bibinfo {volume} {17}},\
  \bibinfo {pages} {053014} (\bibinfo {year} {2015})}\BibitemShut {NoStop}%
\bibitem [{\citenamefont {Cao}(2020)}]{cao2020discrete}%
  \BibitemOpen
  \bibfield  {author} {\bibinfo {author} {\bibfnamefont {Z.}~\bibnamefont
  {Cao}},\ }\bibfield  {title} {\bibinfo {title} {Discrete-phase-randomized
  measurement-device-independent quantum key distribution},\ }\href@noop {}
  {\bibfield  {journal} {\bibinfo  {journal} {Physical Review A}\ }\textbf
  {\bibinfo {volume} {101}},\ \bibinfo {pages} {062325} (\bibinfo {year}
  {2020})}\BibitemShut {NoStop}%
\bibitem [{\citenamefont {Lo}\ \emph {et~al.}(2005)\citenamefont {Lo},
  \citenamefont {Ma},\ and\ \citenamefont {Chen}}]{lo2005decoy}%
  \BibitemOpen
  \bibfield  {author} {\bibinfo {author} {\bibfnamefont {H.-K.}\ \bibnamefont
  {Lo}}, \bibinfo {author} {\bibfnamefont {X.}~\bibnamefont {Ma}},\ and\
  \bibinfo {author} {\bibfnamefont {K.}~\bibnamefont {Chen}},\ }\bibfield
  {title} {\bibinfo {title} {Decoy state quantum key distribution},\
  }\href@noop {} {\bibfield  {journal} {\bibinfo  {journal} {Physical review
  letters}\ }\textbf {\bibinfo {volume} {94}},\ \bibinfo {pages} {230504}
  (\bibinfo {year} {2005})}\BibitemShut {NoStop}%
\bibitem [{\citenamefont {Wang}\ \emph {et~al.}(2023)\citenamefont {Wang},
  \citenamefont {Wang}, \citenamefont {Yin}, \citenamefont {Wang},
  \citenamefont {Lu}, \citenamefont {Chen}, \citenamefont {He}, \citenamefont
  {Guo},\ and\ \citenamefont {Han}}]{wang2023tight}%
  \BibitemOpen
  \bibfield  {author} {\bibinfo {author} {\bibfnamefont {Z.-H.}\ \bibnamefont
  {Wang}}, \bibinfo {author} {\bibfnamefont {R.}~\bibnamefont {Wang}}, \bibinfo
  {author} {\bibfnamefont {Z.-Q.}\ \bibnamefont {Yin}}, \bibinfo {author}
  {\bibfnamefont {S.}~\bibnamefont {Wang}}, \bibinfo {author} {\bibfnamefont
  {F.-Y.}\ \bibnamefont {Lu}}, \bibinfo {author} {\bibfnamefont
  {W.}~\bibnamefont {Chen}}, \bibinfo {author} {\bibfnamefont {D.-Y.}\
  \bibnamefont {He}}, \bibinfo {author} {\bibfnamefont {G.-C.}\ \bibnamefont
  {Guo}},\ and\ \bibinfo {author} {\bibfnamefont {Z.-F.}\ \bibnamefont {Han}},\
  }\bibfield  {title} {\bibinfo {title} {Tight finite-key analysis for
  mode-pairing quantum key distribution},\ }\href@noop {} {\bibfield  {journal}
  {\bibinfo  {journal} {Communications Physics}\ }\textbf {\bibinfo {volume}
  {6}},\ \bibinfo {pages} {265} (\bibinfo {year} {2023})}\BibitemShut {NoStop}%
\bibitem [{\citenamefont {Lu}\ \emph {et~al.}(2024)\citenamefont {Lu},
  \citenamefont {Wang}, \citenamefont {Li},\ and\ \citenamefont
  {Cao}}]{lu2024asymmetric}%
  \BibitemOpen
  \bibfield  {author} {\bibinfo {author} {\bibfnamefont {Z.}~\bibnamefont
  {Lu}}, \bibinfo {author} {\bibfnamefont {G.}~\bibnamefont {Wang}}, \bibinfo
  {author} {\bibfnamefont {C.}~\bibnamefont {Li}},\ and\ \bibinfo {author}
  {\bibfnamefont {Z.}~\bibnamefont {Cao}},\ }\bibfield  {title} {\bibinfo
  {title} {Asymmetric mode-pairing quantum key distribution},\ }\href@noop {}
  {\bibfield  {journal} {\bibinfo  {journal} {Physical Review A}\ }\textbf
  {\bibinfo {volume} {109}},\ \bibinfo {pages} {012401} (\bibinfo {year}
  {2024})}\BibitemShut {NoStop}%
\bibitem [{\citenamefont {Wang}\ \emph {et~al.}(2025)\citenamefont {Wang},
  \citenamefont {Zhou}, \citenamefont {Lu}, \citenamefont {Hao}, \citenamefont
  {Zhao}, \citenamefont {Zhou}, \citenamefont {Li},\ and\ \citenamefont
  {Bao}}]{wang2025asymmetric}%
  \BibitemOpen
  \bibfield  {author} {\bibinfo {author} {\bibfnamefont {H.-T.}\ \bibnamefont
  {Wang}}, \bibinfo {author} {\bibfnamefont {C.}~\bibnamefont {Zhou}}, \bibinfo
  {author} {\bibfnamefont {Y.-F.}\ \bibnamefont {Lu}}, \bibinfo {author}
  {\bibfnamefont {C.-P.}\ \bibnamefont {Hao}}, \bibinfo {author} {\bibfnamefont
  {Y.-M.}\ \bibnamefont {Zhao}}, \bibinfo {author} {\bibfnamefont {Y.-Y.}\
  \bibnamefont {Zhou}}, \bibinfo {author} {\bibfnamefont {H.-W.}\ \bibnamefont
  {Li}},\ and\ \bibinfo {author} {\bibfnamefont {W.-S.}\ \bibnamefont {Bao}},\
  }\bibfield  {title} {\bibinfo {title} {Asymmetric mode-pairing quantum key
  distribution with advantage distillation},\ }\href@noop {} {\bibfield
  {journal} {\bibinfo  {journal} {Chinese Physics B}\ }\textbf {\bibinfo
  {volume} {34}},\ \bibinfo {pages} {040305} (\bibinfo {year}
  {2025})}\BibitemShut {NoStop}%
\bibitem [{\citenamefont {Li}\ \emph {et~al.}(2025)\citenamefont {Li},
  \citenamefont {Dou}, \citenamefont {Xie}, \citenamefont {Kong}, \citenamefont
  {Liu}, \citenamefont {Ma},\ and\ \citenamefont {Tang}}]{li2025asymmetric}%
  \BibitemOpen
  \bibfield  {author} {\bibinfo {author} {\bibfnamefont {Z.}~\bibnamefont
  {Li}}, \bibinfo {author} {\bibfnamefont {T.}~\bibnamefont {Dou}}, \bibinfo
  {author} {\bibfnamefont {Y.}~\bibnamefont {Xie}}, \bibinfo {author}
  {\bibfnamefont {W.}~\bibnamefont {Kong}}, \bibinfo {author} {\bibfnamefont
  {Y.}~\bibnamefont {Liu}}, \bibinfo {author} {\bibfnamefont {H.}~\bibnamefont
  {Ma}},\ and\ \bibinfo {author} {\bibfnamefont {J.}~\bibnamefont {Tang}},\
  }\bibfield  {title} {\bibinfo {title} {Asymmetric protocols for mode pairing
  quantum key distribution with finite-key analysis},\ }\href@noop {}
  {\bibfield  {journal} {\bibinfo  {journal} {Entropy}\ }\textbf {\bibinfo
  {volume} {27}},\ \bibinfo {pages} {737} (\bibinfo {year} {2025})}\BibitemShut
  {NoStop}%
\bibitem [{\citenamefont {Gottesman}\ \emph {et~al.}(2004)\citenamefont
  {Gottesman}, \citenamefont {Lo}, \citenamefont {Lutkenhaus},\ and\
  \citenamefont {Preskill}}]{gottesman2004security}%
  \BibitemOpen
  \bibfield  {author} {\bibinfo {author} {\bibfnamefont {D.}~\bibnamefont
  {Gottesman}}, \bibinfo {author} {\bibfnamefont {H.-K.}\ \bibnamefont {Lo}},
  \bibinfo {author} {\bibfnamefont {N.}~\bibnamefont {Lutkenhaus}},\ and\
  \bibinfo {author} {\bibfnamefont {J.}~\bibnamefont {Preskill}},\ }\bibfield
  {title} {\bibinfo {title} {Security of quantum key distribution with
  imperfect devices},\ }in\ \href@noop {} {\emph {\bibinfo {booktitle}
  {International Symposium onInformation Theory, 2004. ISIT 2004.
  Proceedings.}}}\ (\bibinfo {organization} {IEEE},\ \bibinfo {year} {2004})\
  p.\ \bibinfo {pages} {136}\BibitemShut {NoStop}%
\end{thebibliography}%

\end{document}